%% file: MAXIJ1305.tex
\documentclass[a4paper,fleqn,usenatbib]{mnras}

\usepackage{newtxtext,newtxmath}

\usepackage[T1]{fontenc}
\usepackage{ae,aecompl}

\usepackage{graphicx}	
\usepackage{amsmath}	
\usepackage{times}
\usepackage{color}
\usepackage{float}
\usepackage{rotating}
\usepackage{url}
\usepackage{threeparttable} 

\include{def_bibtex}
\include{definitions}

\title[MAXI J1305-704]{Dynamical confirmation of a stellar-mass black hole in the transient X-ray dipping binary MAXI J1305-704}
\author[ D Mata S\'anchez et al.]{D. Mata S\'anchez$^{1}$\thanks{E-mail: matasanchez.astronomy@gmail.com}, A. Rau$^{2}$, A. \'Alvarez Hern\'andez $^{3,4}$, T. F. J. van Grunsven$^{5,6}$,\newauthor M. A. P. Torres$^{3,4}$, P. G. Jonker$^{5,6}$\\
\\
$^{1}$Jodrell Bank Centre for Astrophysics, Department of Physics and Astronomy, The University of Manchester, M13 9PL, UK\\
$^{2}$Max-Planck Institute for Extraterrestrial Physics, Giessenbachstr. 1, D-85748 Garching, Germany\\
$^{3}$Instituto de Astrof\'isica de Canarias (IAC), E-38205 La Laguna, Tenerife, Spain\\
$^{4}$Departamento de Astrof\'isica, Univ. de La Laguna, E-38206 La Laguna, Tenerife, Spain\\
$^{5}$SRON, Netherlands Institute for Space Research, Sorbonnelaan 2, 3584 CA, Utrecht, The Netherlands\\
$^{6}$Department of Astrophysics/ IMAPP, Radboud University, Heyendaalseweg 135,6525 AJ, Nijmegen, The Netherlands}

\date{Accepted 2021 June 12. Received 2021 June 10; in original form 2021 April 13}

\pubyear{2021}

\begin{document}
\label{firstpage}
\pagerange{\pageref{firstpage}--\pageref{lastpage}}
\maketitle

\begin{abstract}
MAXI J1305-704 has been proposed as a high-inclination candidate black hole X-ray binary in view of its X-ray properties and dipping behaviour during outburst. We present photometric and spectroscopic observations of the source in quiescence that allow us to reveal the ellipsoidal modulation of the companion star and absorption features consistent with those of an early K-type star ($T_{\rm eff}=4610^{+130}_{-160}\, {\rm K}$). The central wavelengths of the absorption lines vary periodically at $P_{\rm orb}=0.394\pm0.004\, {\rm d}$ with an amplitude of $K_2=554\pm 8\, {\rm km \, s^{-1}}$. They imply a mass function for the compact object of $f(M_1)=6.9\pm 0.3\, M_\odot$, confirming its black hole nature. The simultaneous absence of X-ray eclipses and the presence of dips set a conservative range of allowed inclinations $60\, {\rm deg}<i<82\, {\rm deg}$, while modelling of optical light curves further constrain it to $i=72^{+5}_{-8}\, {\rm deg}$. The above parameters together set a black hole mass of $M_1= 8.9_{-1.0}^{+1.6}\, M_{\odot}$ and a companion mass of $M_2= 0.43\pm 0.16\, M_{\odot}$, much lower than that of a dwarf star of the observed spectral type, implying it is evolved. Estimates of the distance to the system ($d=7.5^{+1.8}_{-1.4}\, {\rm kpc}$) and space velocity ($v_{\rm space}=270\pm 60 \, {\rm km\, s^{-1}}$) place it in the Galactic thick disc and favour a significant natal kick during the formation of the BH if the supernova occurred in the Galactic Plane.
\end{abstract}


\section{Introduction}
\label{intro}

Low-mass X-ray binaries (LMXBs) are gravitationally bound stellar systems comprised of a compact object, either a neutron star or a black hole (BH), and a low mass companion star ($\lesssim 1\, M_{\rm \odot}$). Their close orbits lead to Roche Lobe overflow of the companion star, triggering the transfer of mass from the former to the compact object via an accretion disc. Among LMXBs, those known as transient systems experience \textit{outbursts}, epochs where their accretion discs brighten a few orders of magnitude at all wavelengths. During the outburst state they typically exhibit fast and strong variability over timescales of a year before returning to quiescence, with some exceptions where the outburst lasts from only a few days (e.g., V4641 Sagittarii, \citealt{Munoz-Darias2018}) to several decades (e.g. EXO 0748-676, \citealt{Hynes2009}; GRS~1915+105, \citealt{Deegan2009}). 

Within the known population of LMXBs, only a third have been proposed as candidates to harbour stellar-mass BHs (over $60$ systems, see \citealt{Corral-Santana2016} for a review). This preliminary classification is mostly based on  their X-ray properties during the outburst (e.g., \citealt{Belloni2011}). To confirm their BH nature via mass measurements, optical/near-infrared observations during quiescence, when the accretion disc is fainter (i.e. the companion star relative contribution to these electromagnetic bands is higher), are required. The combination of photometric and spectroscopic observations allows us to characterise and trace the orbit of the companion star, ultimately solving the dynamics of the system and obtaining the mass of the compact object. The current number of dynamically confirmed BHs is 18 (see \citealt{Casares2014}; with the recent addition of MAXI J1820+070, \citealt{Torres2019,Torres2020}; and including GX 339-4 but noticing the revision of its mass function, now barely consistent with a low-mass BH, \citealt{Heida2017}). Studies of the remaining candidates are usually hampered by their unfavourable quiescence properties, such as having faint optical/near-infrared counterparts (e.g., MAXI J1659-152, \citealt{Corral-Santana2018}, \citealt{Torres2021}; XTE J1752-223, \citealt{Ratti2012}, \citealt{Lopez2019}) or not detecting the donor star over their bright accretion discs (e.g., Swift J1357.2-0933, see \citealt{Torres2015}, \citealt{MataSanchez2015b}).

We present a photometric and spectroscopic study of the $r'=21.69\pm 0.17$ quiescent counterpart to the black hole candidate MAXI~J1305-704 (hereafter J1305). J1305 was originally identified as a new X-ray transient in 2012 \citep{Sato2012} making use of the \textit{MAXI} instrument on-board the International Space Station (Monitor of All-sky X-ray Image; \citealt{Matsuoka2009}) equipped with the GSC instrument (Gas Slit Camera; \citealt{Mihara2011}). A bright optical counterpart ($g' \sim 16.5$) was soon reported as result of dedicated follow-up observations \citep{Greiner2012}. The combination of the X-ray and optical properties during the outburst led to the classification of this transient as a LMXB, and in particular, it was suggested to harbour a stellar-mass BH (see e.g., \citealt{Greiner2012}, \citealt{Kennea2012}, \citealt{Suwa2012}, \citealt{Morihana2013}). Further observations in the X-ray regime revealed periodic dipping features in its light curve during the outburst (\citealt{Shidatsu2013}). Such dipping behaviour has been previously observed in a handful of LMXBs (e.g., MAXI J1659-152, \citealt{Kennea2011} \citealt{Kuulkers2013}; MAXI J1820+070, \citealt{Kajava2019}; XTE J1710-281, \citealt{Raman2018}), and it is typically explained as obscuration of the X-rays emitted from the central region by dense structures located above the (outer) accretion disc (see e.g., \citealt{White1982}). The dipping behaviour of J1305, combined with the non-detection of eclipses by the companion star and the fact it did not behave as an accretion disc corona source (a thermal component from the accretion disc is present in the X-ray spectrum, \citealt{Shidatsu2013}) suggests a high inclination for the system, but not fully edge-on (see e.g., \citealt{Frank1987}). The periodicity of the dips in J1305 has been associated with the orbital period of the binary ($P_{\rm dip}= 9.74\pm 0.04\,{\rm h}$, \citealt{Shidatsu2013}). However, the complexity of the dipping behaviour (e.g. the presence of both deep and shallow dips) has led to different estimates from other authors ($1.5\, {\rm h}$ or $2.7\,{\rm h}$, \citealt{Kennea2012}; $5\,{\rm h}$, \citealt{Shaw2017}).

While the outburst of J1305 has been studied in detail in the X-rays, its follow-up at other wavelengths has been scarce (e.g., \citealt{Shaw2017} presents the only published optical spectrum obtained during outburst). The non-detection of an optical/near-infrared counterpart in DSS and 2MASS all-sky surveys prior to the discovery outburst \citep{Greiner2012} suggested that J1305 might be too faint for optical quiescent spectroscopy studies. On the contrary, we report in this paper the detection of the quiescent optical counterpart of J1305, and characterise it through both photometric and spectroscopic observations (Sec. \ref{observations}). A careful analysis of both data sets enables us to detect not only an orbital modulation in the light curve, but also to trace the radial velocity of the companion star (Sec. \ref{results}). This allows us to obtain the dynamical solution of the system, including the orbital period as well as confirming the BH nature of the compact object (Sec. \ref{discussion}). The results are summarised in Sec. \ref{conclusion}.

\section{Observations}
\label{observations}

In 2014 we performed photometric observations of J1305 which revealed the quiescence optical counterpart and the characteristic ellipsoidal modulation, which arises from its tidally distorted shape. Based on these initial results, we concluded that the expected relative contribution of the companion star to the optical light might be high enough to detect its absorption features in an optical spectrum. For this reason, we performed a new set of observations with simultaneous photometry and spectroscopy in 2016. Two further (and shorter) blocks of time-resolved photometry were executed two years later (2018) with the aim of analysing the long-term evolution of the system.

\subsection{Photometry}
\begin{figure}
\begin{centering}
\fbox{\includegraphics[width=0.42\textwidth,angle=0]{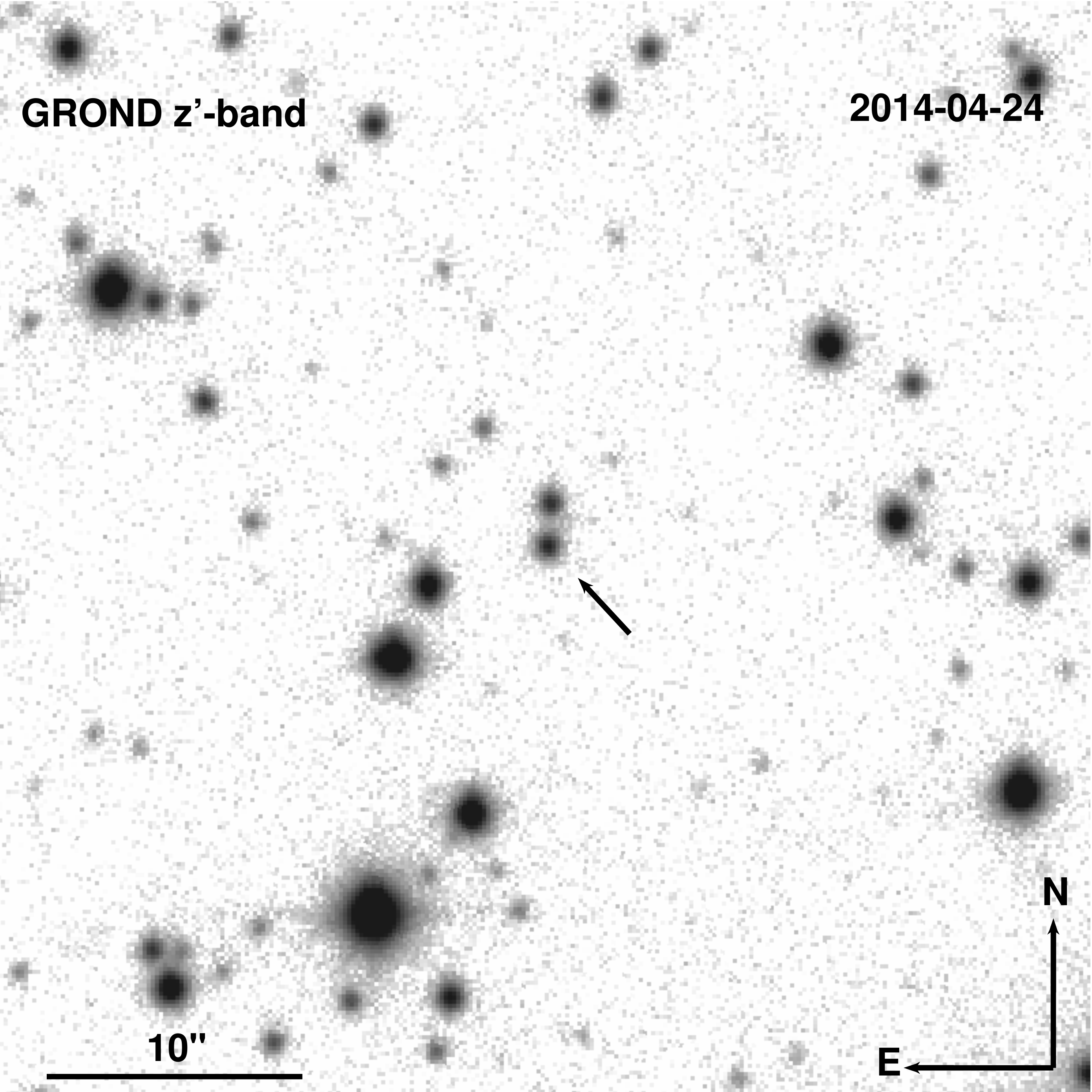}}
\caption{GROND z$^\prime$-band image of the field of J1305 taken on 2014 April 24 (690\,s exposure, 0.7$^{\prime\prime}$ FWHM). The arrow indicates the position of the optical
  counterpart.}
\label{fig:grondFinder}
\end{centering}
\end{figure}

Time resolved imaging was performed using the 7-channel Gamma-Ray burst Optical/Near-infrared Detector  GROND \citep{Greiner2008} mounted at the MPG 2.2\,m  telescope at the ESO La Silla Observatory (Chile). GROND provides simultaneous data in the g$^\prime$, r$^\prime$, i$^\prime$, z$^\prime$, J, H, and K$_s$ bands. We show in Fig.~\ref{fig:grondFinder} a z$^{\prime}$-band finding chart of the field around J1305 approximately two years after the outburst and with the source in quiescence. The analysis reported in this paper is based on the optical (g$^\prime$, r$^\prime$, i$^\prime$, and z$^\prime$) data taken during four nights between April 2014 and June 2018 (see Table~\ref{tab:GROND_lateLog}).

A field star to the North of J1305 (see Fig. \ref{fig:grondFinder}) has been detected also by Gaia DR2 release \citep{GaiaDR22018}, which provides a precise angular separation between them of $1.64''$. This limited our analysis to the optical bands, given the generally low signal-to-noise ratio at near-infrared wavelengths from J1305 and the coarser pixel size of the detectors in this regime ($0.60''/{\rm pixel}$). The 2014 epoch has a consistently low seeing at optical wavelengths, allowing us to confidently isolate the flux of J1305 from that of the nearby star. Photometry of the remaining epochs is also extracted, but we note that partial blending with the contaminant star limits its reliability.

\begin{table}
\begin{center}
\caption{GROND photometry of J1305 in quiescence. We report the observation date, the number of exposures per epoch and the seeing. The mean magnitude in each optical band is provided for the different epochs, where the uncertainties combine that of the statistical errors and the intrinsic variability of J1305.}
\label{tab:GROND_lateLog}
\begin{tabular}{lccc}
\hline
\hline
UTC & \# & Seeing & Mean magnitude\\
& & \\
\hline
2014 Apr 23/24 & 34 & 0.7$^{\prime\prime}$--1$^{\prime\prime}$ & $ {\rm g}^\prime = 21.77\pm 0.12$ \\
&&& $ {\rm r}^\prime = 20.62\pm 0.09$\\
&&& $ {\rm i}^\prime = 20.21\pm 0.08$\\
&&& $ {\rm z}^\prime = 19.93\pm 0.07$\\
2016 Apr 01 $^{{\rm a}}$ & 26  & 1.1$^{\prime\prime}$--2.0$^{\prime\prime}$ & $ {\rm g}^\prime = 21.55\pm 0.11$ \\
&&& $ {\rm r}^\prime = 20.48\pm 0.07$\\
&&& $ {\rm i}^\prime = 20.11\pm 0.06$\\
&&& $ {\rm z}^\prime = 19.86\pm 0.11$\\
2018 June 02/03 & 11 & 0.9$^{\prime\prime}$--1.3$^{\prime\prime}$ & $ {\rm g}^\prime = 21.78\pm 0.10$ \\
&&& $ {\rm r}^\prime = 20.60\pm 0.07$\\
&&& $ {\rm i}^\prime = 20.21\pm 0.08$\\
&&& $ {\rm z}^\prime = 19.93\pm 0.06$\\
2018 June 14/15 & 15 & 1.0$^{\prime\prime}$--1.6$^{\prime\prime}$& $ {\rm g}^\prime = 21.59\pm 0.05$ \\
&&& $ {\rm r}^\prime = 20.49\pm 0.04$\\
&&& $ {\rm i}^\prime = 20.10\pm 0.03$\\
&&& $ {\rm z}^\prime = 19.85\pm 0.04$\\
\hline
\end{tabular}
\end{center}
\raggedright
$^{\rm a}$: simultaneous with VLT/FORS2 spectroscopy (see \S~\ref{sec:forsdata})
\end{table}

In total, 86 observations with exposures of 690\,s each were obtained in each band. The data were reduced and analyzed with the standard tools and methods described in \cite{Kruehler:2008aa}. The photometry was  obtained using point spread function fitting and calibrated relative to a reference image in each band taken under photometric conditions. The absolute photometric calibration in the AB-magnitude system was derived using the zero points from an observation of a field covered by the SDSS Data Release 8 \citep{Aihara:2011aa} taken shortly after the reference images. The resulting 1$\sigma$  systematic uncertainties   are   0.04\,mag  in g$^\prime$, r$^\prime$, i$^\prime$ and z$^\prime$.

\subsection{Spectroscopy}
\label{sec:forsdata}

We observed J1305 with the Very Large Telescope Unit Telescope 1 (VLT-UT1; Paranal Observatory, Chile) equipped with the FOcal Reducer/low dispersion Spectrograph 2 (FORS2, \citealt{Appenzeller1998}) in long-slit mode. Our setup employs the GG435 filter and the 1200R grism combined with a slit of width $1.0''$ and the CCD detector pixels were read out with a $2\times 2$ binning. This allows us to cover the spectral range of $5750 - 7310\,\text{\AA}$ with a dispersion of $0.76\,\text{\AA}/\rm{pix}$ and a spectral resolution at the central wavelength ($\rm 6530\, \text{\AA}$) of $\rm R\sim 2140$. The spectra were reduced using the {\it EsoReflex} automated data reduction workflow \citep{Freudling2013}. In addition, and given that the wavelength calibration relies on a single arc obtained at the end of the night, we employed the sky line [\ion{O}{I}]=6300.304 \AA $\,$ to correct the wavelength calibration for velocity drifts caused by potential flexure effects.

Our final spectroscopic data set consists of 16 spectra (covering $5800-7300$ \AA) of 1800s exposure time each (except the last spectrum, which was only integrated during 1200s), consecutively obtained during a single observing run (2016 March 31, simultaneous with 2016 photometric epoch) and lasting $\sim 9\, {\rm h}$ (i.e. almost covering the longest $0.4\, {\rm d}$ period established from X-ray dips). All the spectra have been corrected to the barycentric reference frame using \textsc{astropy} utilities with the built-in ephemeris, and their observation times set to that of mid-exposure (Barycentric Julian Date, BJD). The seeing during the spectroscopic observations was $0.6-1.0''$, measured as the full-width-at-half-maximum (FWHM) of a Gaussian fit to the target profile after collapsing in the wavelength dimension. 

We asses a potential contamination of the J1305 spectra by the nearby field star. The slit orientation was set to follow the East-West direction, therefore perpendicular to the line connecting the target and the nearby star. Acquisition images were obtained every three hours to help at correcting any misplacement of the slit on the target. Under the assumption of the slit being perfectly centred on the target, and given the nearby star relative brightness (half the flux of J1305 at all times), we conclude that the potential contamination of the optical spectra should be negligible (< 10\% of the total flux). Nevertheless, any departure from the aforementioned conditions could increase the interloper fractional contribution (see a further discussion on this topic in Sec. \ref{discveil}).

\section{Results}
\label{results}

The distorted shape of the Roche Lobe filling secondary produces a variable quiescence light curve at the orbital period of the binary. In addition, the companion star detection enables dynamical studies on these binaries. On the contrary to other accreting binaries, such as transitional millisecond pulsars, \citep{Archibald2009} or low-state cataclysmic variables \citep{RodriguezGil2015}, the accretion disc responsible for the outburst behaviour of LMXBs does not completely disappear during quiescence (as proven by the presence of broad hydrogen emission lines in their optical spectra). The contribution of the accretion disc to the quiescent spectrum continuum effectively makes the companion star absorption lines to appear shallower, in the most extreme cases rendering them undetectable (e.g., \citealt{MataSanchez2015b, Torres2015}).

\subsection{Optical spectroscopy}

The lack of a reliable flux calibration of the spectra led us to divide each of them by a low-order polynomial fit of the continuum. We will refer hereafter to them as normalised spectra.

\subsubsection{The radial velocity curve}
\label{rvcurve}

We search for the companion star absorption features by analysing the cross-correlation functions (CCFs) resulting from the \textit{crosscorrRV} \textsc{python} routine from PyAstronomy\footnote{https://github.com/sczesla/PyAstronomy}. This routine allows us to compare each observed spectrum with stellar templates from a synthetic grid generated by \citet{Coelho2014}. The grid of templates, covering effective temperatures $T_{\rm eff}=4000-9000\, {\rm K}$ with a spectral resolution of $15\,  {\rm km\,s^{-1}}$ and limited to dwarf stars of solar metallicity, was re-binned and broadened via a Gaussian convolution to a spectral resolution matching that of our observed FORS2 spectra ($\rm 140\, km\,s^{-1}$). We focus our analysis on the spectral range $\sim 6000-7000$ {\AA}, where prominent features for late-type stars are expected. We mask out emission lines arising from the accretion disc, telluric absorption bands, and wavelength ranges contaminated by strong sky-subtraction residuals. We obtain the CCFs by comparing each observed spectrum with the selected template, inspecting a range of velocity shifts of -1000 to 1000 $\rm km \, s^{-1}$ in steps of $5\, \rm km \, s^{-1}$.

We find consistent CCFs for the temperature range of $T_{\rm eff}=4000-7000\, {\rm K}$, while higher temperature templates yield unreliable results because their spectral features are too different from those observed in J1305. This is an expected outcome as we cannot employ in the CCFs hydrogen lines (which are present over a wider range of spectral types) due to the dominant contribution of the disc to the line profiles. As a sanity check on our procedure, we also inspected the CCFs resulting from comparison with observational templates from the INDO-US stellar library (spectral resolution of $\rm  50\, km\,s^{-1}$, \citealt{Valdes2004}) after broadening them to the resolution of our data, and found fully consistent results. We will discuss hereafter the results from the comparison with the synthetic template of $T_{\rm eff}=4500\, {\rm K}$, corresponding to the companion star effective temperature derived in Sec. \ref{spec_class}.

\begin{figure}
\includegraphics[keepaspectratio, trim=0cm 0cm 0cm 0cm, clip=true, width=0.5\textwidth]{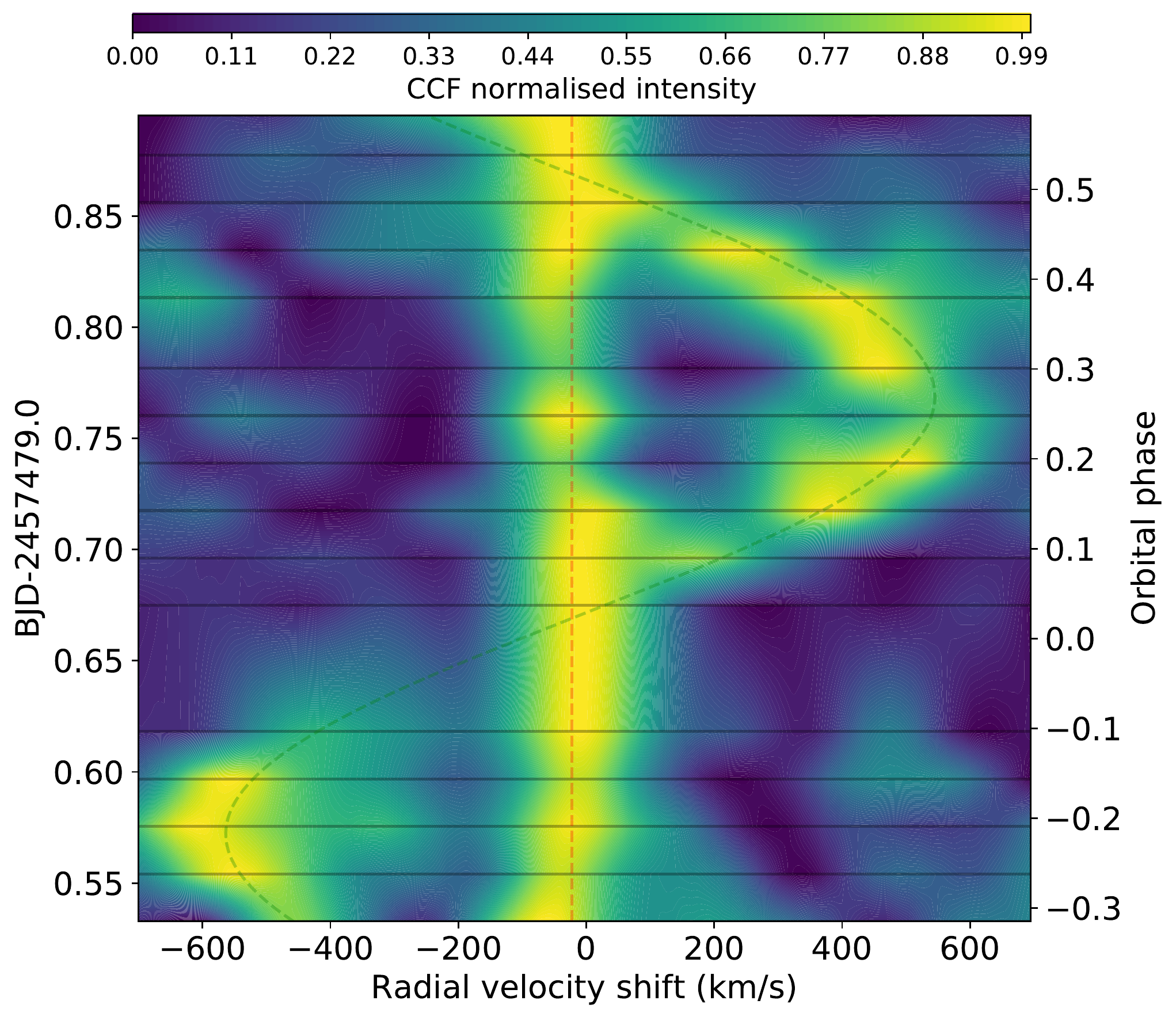} \caption{Trail of the CCFs resulting from the comparison of each individual spectrum of J1305 with a $T_{\rm eff}=4500\, {\rm K}$ stellar template. The x-axis defines the radial velocity shift applied to the template, while the y-axis corresponds to the BJD of each spectrum (left-hand axis) and orbital phase (right-hand axis, adopting the spectroscopic orbital period derived in this work). We mark as horizontal, black lines the mid-exposure times of our observations. The color map depicts the intensity of the CCFs and has been normalised between the maximum and minimum values for each CCF separately for display purposes. The best sinusoidal fit of the data has been overplotted as dashed green line for peak $B$, while the mean value for peak $A$ is plotted as a dashed red line.} 
\label{xcor}
\end{figure}

Absorption features in the spectra originating in the companion star in an X-ray binary are expected to produce a single peak in the CCF, where its peak velocity corresponds to the required shift to best match the template. Therefore, they are expected to vary as a function of orbital phase. This variation can be used to construct the radial velocity curve of the companion star. However, in our data we find two clear features  in the trail of CCFs shown in Fig. \ref{xcor}: an apparently stationary peak (peak $A$), and a peak of comparable intensity to the former whose centroid radial velocity varies smoothly with time (hereafter peak $B$).

\begin{figure*}
\includegraphics[keepaspectratio, trim=0cm 0cm 0cm 0cm, clip=true, width=\textwidth]{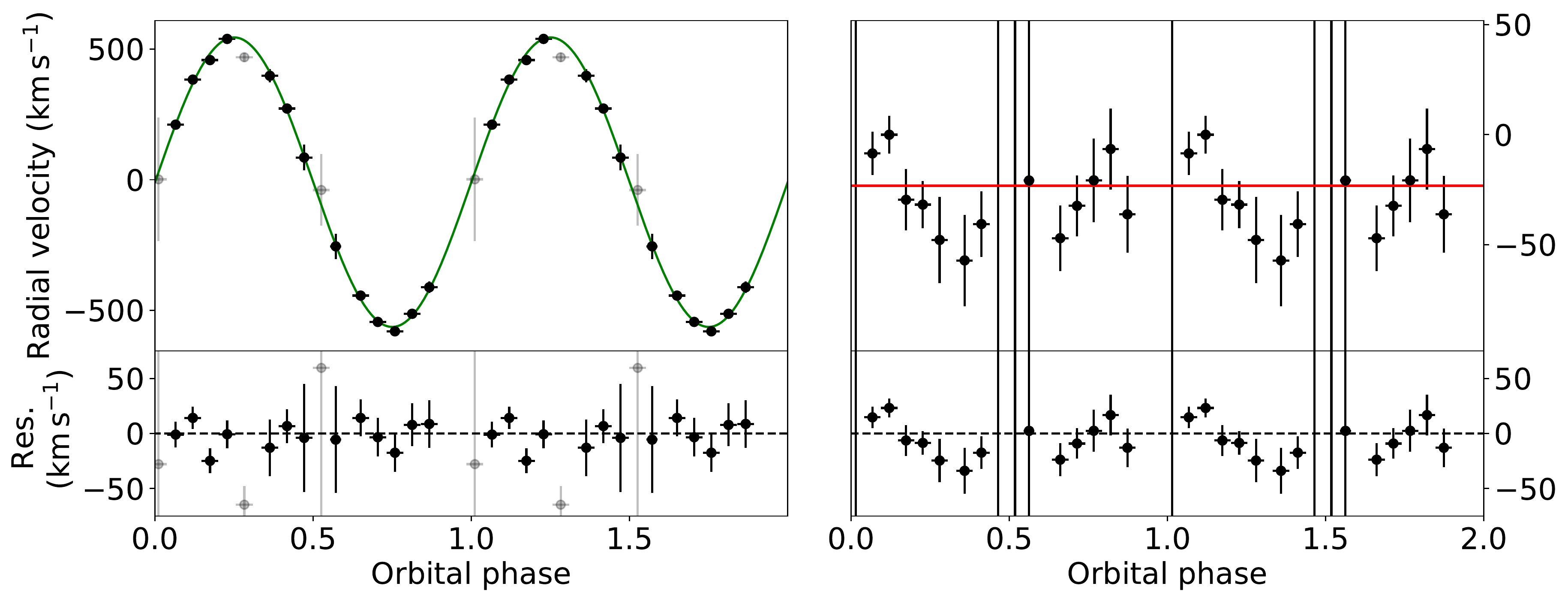} \caption{Phase-folded radial velocity curves from the two CCF components following the ephemeris calculated in Sec. \ref{rvcurve}. The orbital cycle is repeated twice for the sake of clarity. Left figure, top panel: Peak $B$ radial velocity curve plotted as black filled circles. The vertical error bars correspond to those derived in the MC analysis, and the length of the horizontal bars indicate the exposure time of each spectrum divided by the best-fit period. We show in grey points the main outlier velocity and all points with error bars too large to contribute significantly to the sinusoidal fit, that is displayed with a solid, green curve. Right figure, top panel: Radial velocity curve for peak $A$, marked as black filled circles, with the mean value of the sample marked as a solid red line. Bottom panels: Residuals of each fit.}
\label{rvcurvefig}
\end{figure*}

In order to analyse the double peaked CCFs, we employ the \textit{minimize} \textsc{python} routine to fit each CCF with a model consisting of two Gaussian components, both of them having the same FWHM. The resulting Gaussian centroids from this fit allow us to construct radial velocity curves for each of the two peaks. The uncertainties on the radial velocities derived through this method are purely statistical, and probably underestimate the real value. To obtain more realistic uncertainties, we perform a Monte Carlo analysis. We simulate 1000 spectra from each observed spectrum using the real data as a seed and assuming that both their wavelength and flux values follow Gaussian distributions with standard deviations given by the root-mean-square of the wavelength calibration and the flux uncertainty, respectively. We repeat the cross-correlation described above to each simulated spectrum, obtaining a distribution of radial velocities associated with each observed spectrum. We will employ the mean and standard deviation drawn from these distributions to construct the final radial velocity curve for peaks $A$ and $B$ of the CCF. From their analysis, it follows that:
\begin{itemize}

\item The radial velocities from peak $B$ exhibit a clear periodic modulation. Hence we fit them to a sinusoid of the form:
$$v(t)=\gamma_B+K_B\,\sin{\dfrac{2\pi (t-T_{B})}{P_{B}}} $$
where $v(t)$ is the measured peak velocity, $t$ is the observation BJD, $T_{B}$ is the reference time for phase zero, $P_{B}$ is the period of the signal and $\gamma_B$ allows for an offset in radial velocity. The resulting fit gives $\chi^{2}_{\rm red}=19.98/12\, {\rm d.o.f.}$ (where the degrees of freedom are defined as the number of datapoints minus the number of parameters). The best fit parameters are $K_B= 546\pm 9 \,{\rm  km \, s^{-1}};\, \gamma_B= -13\pm 6 \,{\rm  km \, s^{-1}};\, P_B= 0.395\pm 0.006 \,{\rm d}; \,  T_B= 2457479.6695\pm 0.0015 {\rm d}$, 
where uncertainties are quoted at $\rm 1\sigma$. Discarding a single outlier of the radial-velocity curve as well as two other points with high error bars (see Fig. \ref{rvcurvefig}), improves the fit significantly ($\chi^{2}_{\rm red}=9.33/9\, {\rm d.o.f.}$) and gives consistent parameters that will be hereafter employed:

\begin{align*}
     K_B= 554\pm 8 \,{\rm  km \, s^{-1}};\qquad \gamma_B= -9\pm 5 \,{\rm  km \, s^{-1}}
\end{align*}
\begin{align*}
 P_B= 0.394\pm 0.004 \,{\rm d}; \qquad  T_B ({\rm BJD})= 2457479.6705\pm 0.0013
\end{align*}

The resulting period is close to the that measured from X-ray dips ($P_{\rm dip}=0.4058\pm 0.0017\,{\rm d}$, \citealt{Shidatsu2013}), and matches that found in the photometric modulation (see Sec.\ \ref{photometrylc}). Given the coincidence with the above periodicities, we conclude that the companion star photosphere is the origin of the absorption features producing this signal in peak $B$. In this scenario, the physical interpretation of these parameters is straightforward: $K_B$ is the radial velocity semi-amplitude of the companion star ($K_2$), $\gamma_B$ is the systemic radial velocity ($\gamma$), $T_B$ is the BJD for the zero phase which corresponds to the companion star inferior conjunction ($T_0$) and $P_B$ is the orbital period of the binary ($P_{\rm spec}$). At this point, it is worth inspecting the correlation between $K_2$ and the FWHM of the $\rm H\alpha$ line found by \citet{Casares2015} for quiescence LMXBs. In order to account for variability in the emission line profile, we use the FWHM of $\rm H\alpha$ measured as the mean and standard deviation from a single Gaussian fit to the individual spectra: ${\rm FWHM}=2450 \pm  200 \, {\rm km\, s^{-1}}$. The correlation between $\rm FWHM$ and $K_2$ predicts $K_2=570\pm 60  \, {\rm km\, s^{-1}}$. In order to avoid a potential overestimation of the FWHM by a narrow absorption line component in the $\rm H\alpha$ profile unrelated to J1305 (see Sec. \ref{discveil}), we followed \citet{Torres2019a} and masked out the core of the line (between $-79$ and $61\, \rm{km\, s^{-1}}$, centred at the systemic velocity). This results in a ${\rm FWHM}=2350 \pm  180 \, {\rm km\, s^{-1}}$ and therefore $K_2=550\pm 50  \, {\rm km\, s^{-1}}$. Both results are fully consistent with the value determined from the radial velocity curve, strengthening the association of peak $B$ in the CCF with the companion star. 

\item A visual inspection of peak $A$ radial velocity curve (see Fig. \ref{rvcurvefig}) suggests a potential low-amplitude modulation at the same period determined for peak $B$ (Fig. \ref{rvcurvefig}). To account for this possibility, we performed a sinusoidal fit where the period is fixed to $P_B$, resulting in $\chi^{2}_{\rm red}=9.06/13\, {\rm d.o.f.}$ and best fit parameters: $K_A=27\pm 6 \, {\rm km\, s^{-1}}$, $\gamma_A = -29\pm 4 \, {\rm km\, s^{-1}}$ and $T_A ({\rm BJD})= 2457479.98\pm 0.03$. However, the low $\chi^{2}$ of the fit warns about a possible overfit to the data. The inconsistency between $\gamma_A$ and $\gamma_B$ does not favour an association with J1305 either. Analysis of the sky line [\ion{O}{I}] 6300.304 \AA $\,$ during the spectra reduction showed its centroid radial velocity smoothly drifted from the laboratory rest frame as a consequence of our observations being condensed into a single night and the wavelength calibration depending on a single arc. While the spectra has already been corrected in first order from the measured velocity shifts, a residual effect might still remain and generate an artificial low-amplitude variability in the otherwise stationary features. For this reason, we dismiss the potential variability of peak $A$ and consider it hereafter as a stationary set of absorption lines, described by its mean value and standard deviation: $\gamma_A = -30\pm 17 \,{\rm  km \, s^{-1}}$.

\end{itemize}

\subsubsection{Spectral classification and rotational broadening of the companion star}
\label{spec_class}

\begin{figure*}
\includegraphics[keepaspectratio, trim=0cm 0cm 0cm 0cm, clip=true, width=\textwidth]{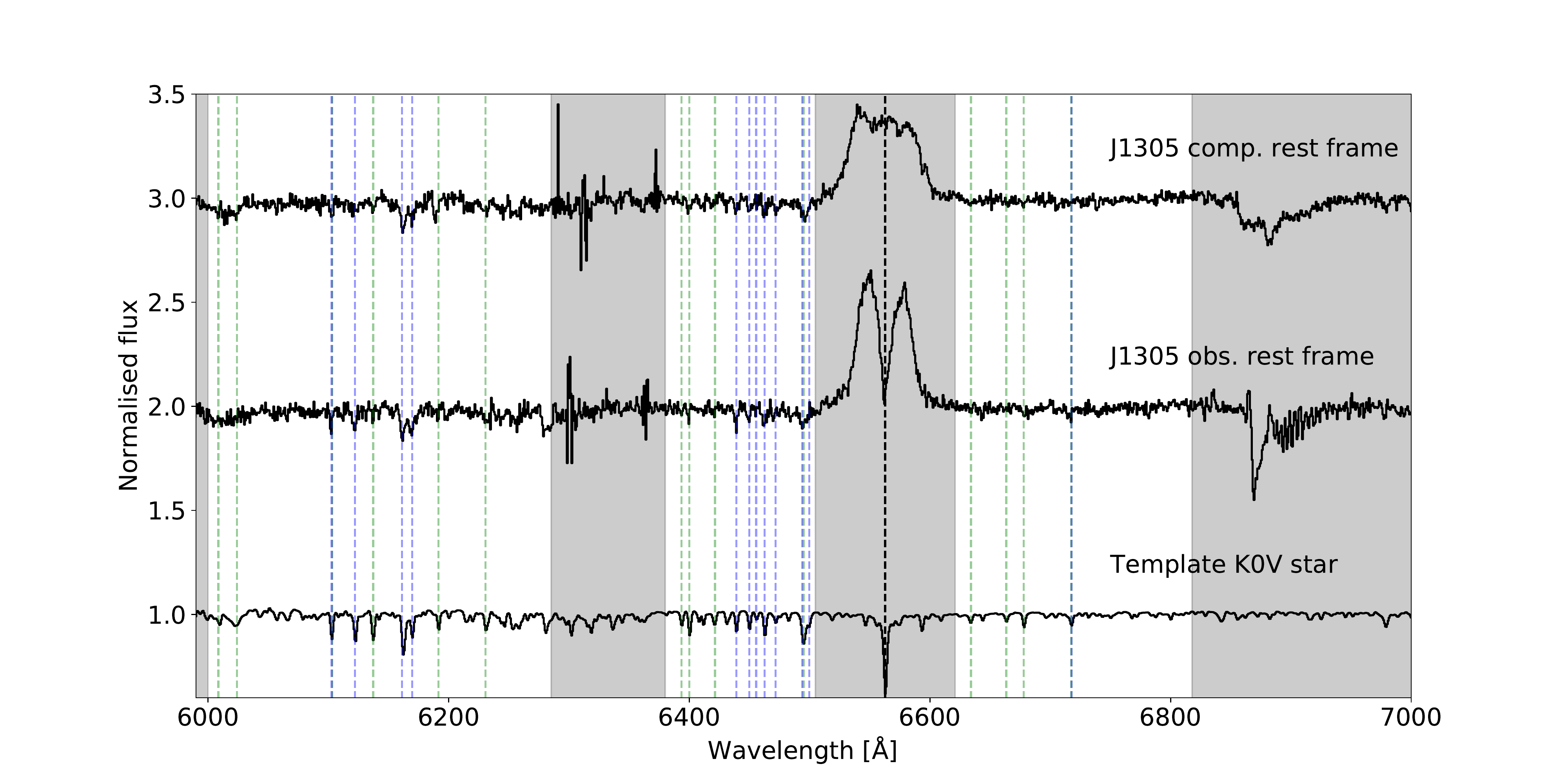} \caption{From top to bottom: Normalised averaged spectrum of J1305 in the companion star rest frame, normalised averaged spectrum of J1305 in the observer rest frame, and a K0V star template (from \citealt{Valdes2004}) broadened to the VLT/FORS2 spectral resolution. The spectra have been offset in the vertical axis for display purposes. Dashed lines mark the $\rm H\alpha$ line rest wavelength (black), as well as absorption features characteristic of late-type stars (\ion{Fe}{i} in green and \ion{Ca}{i} in blue). Shadowed areas show the mask applied for both the CCFs and the spectral classification.}
\label{av_spec}
\end{figure*}

We show in Fig.\ \ref{av_spec} the normalised averaged spectrum of J1305 in two different reference frames: the rest frame of the companion star and that of the observer. They are obtained correcting the individual spectra for the measured velocities of the companion star (i.e., peak $B$ of the CCFs) and without applying any velocity shift, respectively. Then, a weighted average using the measured signal-to-noise in a clean region of the continuum ($6700-6800\,{\text\AA}$) is performed. The averaged spectrum in the companion star inertial frame shows absorption features similar to those observed in late-type stars, as proven by the comparison with the K0V template spectrum in Fig. \ref{av_spec}. On the other hand, the averaged spectrum in the observer rest frame does not differ much from that obtained in the compact object reference frame, as BHs in LMXBs have relatively small radial velocities semi-amplitudes ($K_1$) due to their small mass ratios ($q=M_2/M_1=K_1/K_2$), with typical values of $q\sim 0.1$ for a BH mass of $M_1\sim 10\, M_\odot$ and a companion star mass of $M_2\sim 1\, M_\odot$. The averaged spectrum in this inertial frame exhibits, apart from the emission lines from the accretion disc, absorption features similar to those usually found in late-type stars (and to those found in the companion rest frame). In spite of the resemblance between the absorption features of these two averaged spectra, they are showing completely different components, as they are the underlying origin of the two peaks found in the CCFs. As a matter of fact, stationary features shown in the observer rest frame spectrum such as the telluric band at $\sim 6900$ {\AA} or the narrow absorption core component in the double-peaked $\rm H\alpha$ emission line, are heavily smeared in the companion rest frame spectrum as a result of corrections for radial velocities as high as $\pm 10 \, ${\AA} for the derived $K_2$. We discard the origin of peak $A$ stationary lines to be telluric, interstellar or resulting from reduction artifacts, as they appear in clean regions of the spectra. The different systemic velocity respect that of peak $B$ (associated with the companion star) suggests an independent origin from J1305. We propose they arise from contamination of the observed spectra by an interloper star.

We perform the spectral classification of the companion star by inspecting the normalised averaged spectrum in its reference frame. We employ a \textsc{python} script based on \textsc{emcee}, an implementation of a Markov Chain Monte Carlo (MCMC) sampler (\citealt{Foreman-Mackey2013}). We define the likelihood function as the $\chi^2$ resulting from the comparison of our averaged spectrum with the generated models from a grid of templates. Here we employ a synthetic grid of template star spectra from \citet{Coelho2014}, and restrict our analysis to the sub-set of solar metallicity models. To generate a model for any given combination of surface gravity ($\log g$) and effective temperature ($T_{\rm eff}$), we normalise the synthetic spectra and perform a linear interpolation on the spectral grid, which originally covered $T_{\rm eff}=4000-9000\, {\rm K}$ mostly uniformly in steps of $250\, {\rm K}$, and $\log g = 3.5 - 4.5$ in steps of $0.5$. Then, this model is convolved with a Gaussian kernel to match the target spectral resolution. The effect of the stellar rotation on the broadening of the photospheric lines is implemented through the \textsc{Pyastronomy} routine \textsc{fastRotBroad}, which follows the prescription described in \citet{Gray1992} and allows for a linear limb-darkening law. We simulate the effect of the veiling of the absorption lines by the accretion disc adopting a contribution to the continuum flux constant at all wavelengths. For this purpose, we define the veiling factor (at the observed range of wavelengths, covering r$^\prime$ band) as the ratio of fluxes of non-stellar origin to the total emitted light: $X=F_{\rm non-stellar}/F_{\rm tot}$. The normalised flux of the model is re-scaled as follows: $X+(1-X)f_{\rm model}$.

We explore a range of $\log g= 3.5-4.5$ and $T_{\rm eff}=4000-9000\, {\rm K}$, assuming uniform priors for both of them between these limits. We also assume uniform priors on the veiling factor ($X=0-1$) and the rotational broadening projected in the line of sight ($v \sin i =20- 200 \, {\rm km\, s^{-1}}$, were we adopt a linear limb darkening $\epsilon=0.7$, \citealt{AlNaimiy1978}). We also fix the instrumental broadening to a value corresponding to our FORS2 observations ($140\, {\rm km\, s^{-1}}$).

\begin{figure*}
\includegraphics[keepaspectratio, trim=0cm 0cm 0cm 0cm, clip=true, width=\textwidth]{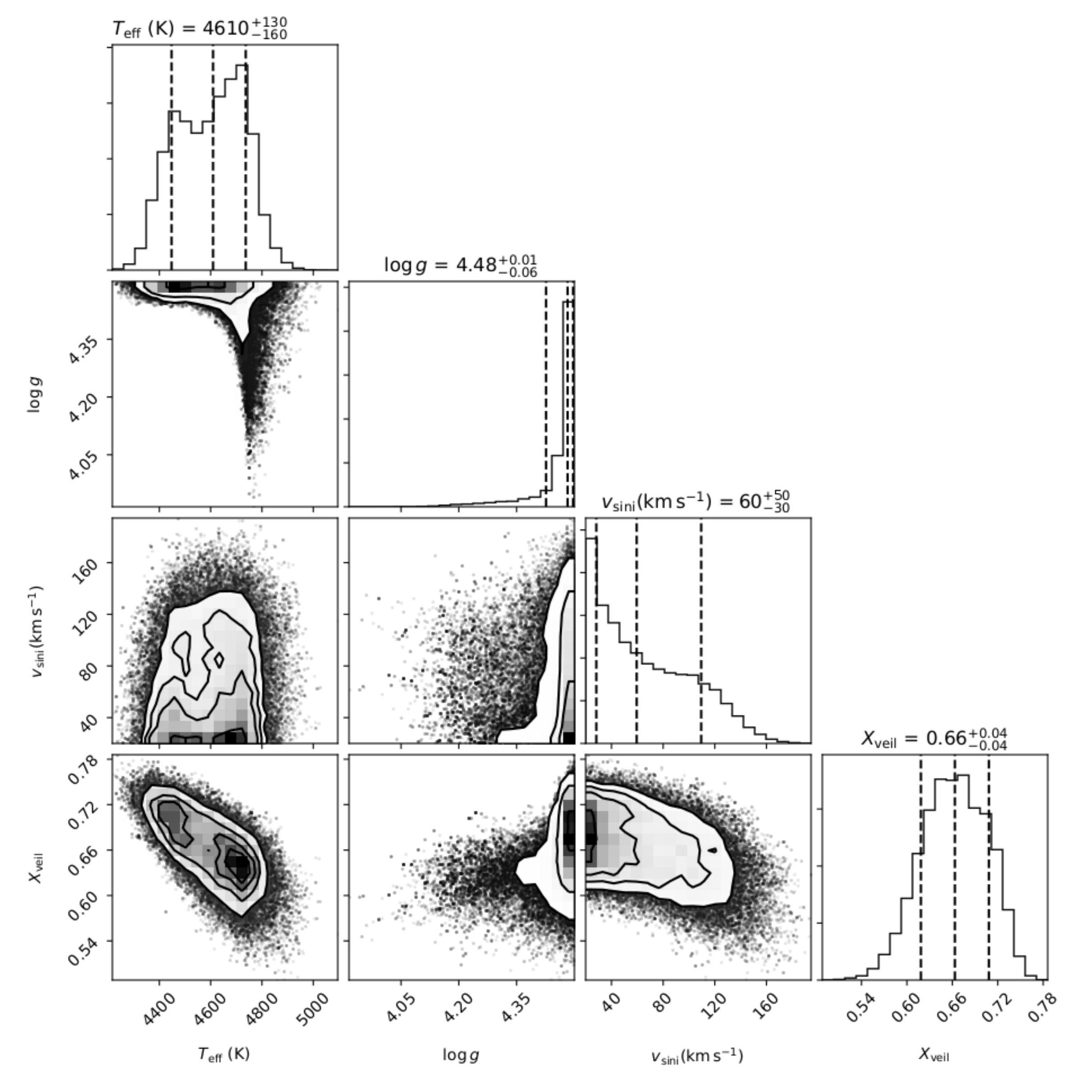} \caption{Corner plot resulting from the spectral classification of J1305's averaged spectrum in the companion star rest frame through comparison with a grid of synthetic templates. Contours in the 2D plots correspond to the $0.5\sigma$, $1\sigma$, $1.5\sigma$ and $2\sigma$ levels (respectively containing 11.8, 39.3, 67.5 and 86.5 \% of the volume), while in the marginalised 1-D distributions the 0.16, 0.50 and 0.84 quantiles are marked with dashed lines.}
\label{cornerplot}
\end{figure*}

The corner plot with the MCMC results is shown in Fig \ref{cornerplot}. We report the 16, 50 and 84$\%$ quantiles (also referred as confidence levels, c.l.) of the marginalised distributions for the key parameters of J1305 (which for a Gaussian-like distribution would correspond to the $\rm 1 \sigma$ uncertainty):

\begin{align*}
    T_{\rm eff}=4610^{+130}_{-160}\, {\rm K};\qquad X=0.66\pm 0.04
\end{align*}
\begin{align*}
    v \sin i <110\, {\rm km\, s^{-1}};\qquad \log g>4.42
\end{align*}

The median values of the effective temperature and the veiling factor are sensible representatives of the obtained distributions, where the departure from a Gaussian-like shape is well described by the asymmetric uncertainties. The surface gravity distribution clusters near the upper limit of $4.5$ imposed by the priors, favouring $\log g>4.42$. While this is not enough to set strong constraints on the parameter, it clearly disfavours templates of much lower surface gravity than typical dwarf stars. Fig.\ \ref{cornerplot} also shows an anti-correlation between the veiling factor and $T_{\rm eff}$ and $v \sin{i}$, which is as expected. On one side, a higher effective temperature implies shallower metallic lines in the spectrum. On the other side, a higher rotational velocity broadens the lines, again making them shallower. Higher veiling factors imply more diluted (shallower) absorption lines, leading to the anti-correlation with both parameters shown in the corner plot.

The component of the rotational velocity projected onto the line of sight ($v \sin i$, where $i$ is the orbital inclination) in combination with $K_2$ provides an independent determination of $q$ through the expression \citep{Wade1988}: $v \sin i \approx 0.462\, K_2\, q^{1/3}\, (1+q)^{2/3}$. The measurement of $v \sin i$ from observed spectra is traditionally performed through the optimal subtraction technique (e.g., \citealt{Marsh1994}). The spectral classification method implemented above is essentially constructed around this technique, but allows us to explore a larger parameter space simultaneously. However, given the low resolution of our FORS2 data, our analysis shows a non-Gaussian distribution for this parameter that rams against the minimum values allowed. As a complementary test, we compare the averaged spectrum in the companion star rest frame with a single synthetic template of $T_{\rm eff}=4500\, {\rm K}$ and $\log g=4.5$ (the closest template from the original grid to the spectral classification obtained here) following the MC approach described in \citet{Torres2020} (see also \citealt{Jonker2007}). This method relies on subtracting from the weighted average target spectrum a broadened version (with a rotational profile as defined in \citealt{Gray1992}) of the template scaled by a factor between 0 and 1 to account for the accretion disc veiling. Then, we compare the $\chi^2$ of the residuals with a smoothed version of themselves. After inspection of different FWHMs for the Gaussian employed to smooth the residuals, we conclude that the best results correspond to $\rm{FWHM}=50\,{\text\AA}$. The distribution of $v \sin i$ obtained with this technique is bi-modal, with a first peak consistent with null velocity and a second, broader peak centred at $ 62\pm 17\, {\rm km\, s^{-1}}$ (c.l. interval of 16-84\% ). These results are consistent with those obtained from the spectral classification, and lead us to conclude that the spectral resolution of our data is not high enough to evaluate $v \sin i$ for J1305. We will consider hereafter the upper limit derived from the spectral classification, $v \sin i <110\, {\rm km\, s^{-1}}$ ($84$ per cent c.l.).

To test the reliability of our results, we have performed further \textsc{emcee} tests, such as leaving the limb darkening as a free parameter, or applying a Gaussian prior on the instrumental resolution of $140\pm 10\, {\rm km\, s^{-1}}$ (in an attempt to account for the variability in spectral resolution across the spectral range). All the results are fully consistent with those presented above, and they do not provide further insight on the parameters of the system.

\subsection{Light curve modelling}
\label{photometrylc}

We first attempt to measure the orbital period through a Lomb-Scargle periodogram performed on the light curves obtained in 2014 and 2016, which have the longest phase coverage. The periodogram produces a broad peak at frequency $\nu = 5.4\pm 0.9\, {\rm d^{-1}}$, where the uncertainty corresponds to the standard deviation of a Gaussian fit to the peak in the Lomb-Scargle periodogram. If the ellipsoidal modulation is the dominant periodic component of the light curve, we would expect the orbital period to be twice the value obtained from the periodogram. This implies a photometric period of $ 0.37\pm 0.06\, {\rm d}$, a value fully consistent with the results from the spectroscopic analysis ($P_{\rm spec}=0.394\pm 0.004\, {\rm d}$).

We continue our analysis by modelling only the 2014 epoch of observations, which provides us with a fully sampled orbital cycle under the best seeing conditions (Tab. \ref{tab:GROND_lateLog}). We combine the light curve modelling code \textsc{XRbinary}\footnote{The  full description of the \textsc{XRbinary} code is available at: http://www.as.utexas.edu/~elr/Robinson/XRbinary.pdf} (see \citealt{Bayless2010}) and the \textsc{emcee} sampler \citep{Foreman-Mackey2013} to fit the photometric data of J1305. The  \textsc{XRbinary} code allows us to simulate the contribution to the light curve from different binary components. First and foremost, the companion star (assumed to be Roche Lobe filling and tidally locked) is responsible for the ellipsoidal modulation. In addition, as supported by the spectroscopic analysis, the flux contribution of an accretion disc must be included. A single bright spot at the edge of the disc will be introduced to reproduce departures from the ellipsoidal modulation. 

The \textsc{XRbinary} models are defined by the following parameters: the binary inclination $i$, photometric orbital period ($P_{\rm phot}$), phase offset ($\phi_0$, defined with respect to $T_{0} ({\rm BJD})=2456771.6463$, an estimated value based on visual inspection), $q$, $K_2$ and the companion effective temperature (we will label the latter as $T_2$ to distinguish it from that determined from spectroscopy, $T_{\rm eff}$). We consider a cylindrically symmetric accretion disc with a height  profile defined by:
$$h (r) = H_{\rm d} \left( \dfrac{r-r_{\rm in}}{R_{\rm d}-r_{\rm in}}\right)^{n} \qquad r_{\rm in} \leq r \leq R_{\rm d}$$
where $r_{\rm in}$ and $R_{\rm d}$ are the inner and outer disc radii; $H_{\rm d}$ is the semiheight at the disc edge and the exponent $n$ defines the height profile. The temperature profile of the disc corresponds to that  of a steady-state, optically thick, viscous disc:
$$T^4(r) = \dfrac{K}{r^3}\left(1-\left(\dfrac{r_{\rm in}}{\rm r}\right)^{1/2}\right) \qquad r_{\rm in} \leq r \leq R_{\rm d}$$ 
where the normalisation constant $K$ allows  the temperature distribution yield the bolometric luminosity $L_{\rm d}$ of the disc. $T_{\rm edge}$ defines the temperature of the outer edge of the disc. Finally, we include a bright spot as described in \citet{vanGrunsven2017}; fixed at the disc edge, expanding across the full disc height ($2 \, H_d$) and defined by its temperature ($T_{\rm spot}$), azimuth angle ($\zeta$, which equals zero in the direction from the companion star to the compact object, and increases in the direction opposite to the orbital motion of the star), and width ($\zeta_{\rm width}$). This can be associated with a hot-spot, a region where the accretion stream from the companion impacts on the accretion disc.

Many of the accretion disc parameters have little influence on the final fit (in part due to the limited signal-to-noise of our observations), and will be hereafter fixed to the values compiled in Tab. \ref{tab:lcmodelfixed} following \citet{vanGrunsven2017}. This reduces the free accretion disc parameters to $R_{\rm d}$ and $L_{\rm d}$. In order to further reduce the number of free parameters in our model, we fix $K_2$ to the value measured within this work (Sec. \ref{rvcurve}).

For the following parameters, we consider flat priors to constrain them to physically sensible values: phase offset $\phi_{\rm 0}$ (-0.15 - 0.15), $\zeta_{\rm width}$ ($0 - 180 {\rm deg}$) and $T_2$ ($3000-8000\, {\rm K}$). We set a prior on the orbital inclination assuming an isotropic distribution (i.e. uniform in $\cos{i}$). We also set the mass ratio to be $q>0.01$, a sensible lower limit from the physical point of view, as most of the known BH LMXBs do not show such extreme values (see \citealt{Casares2014}). We constrain $R_d\leq 0.6\, {a}$, where $a$ is the orbital separation between the stellar components. The upper limit was set to the tidal truncation radius at the minimum considered $q$ ($R_T=0.9\, R_{1}$, being $R_{1}$ the volume averaged Roche Lobe radius for the compact object; \citealt{Eggleton1983}, \citealt{Whitehurst1991}). The veiling factor derived in Sec. \ref{spec_class} was used to set a prior on the disc to total flux ratio in the r$^\prime$ band, which bandpass falls within the spectral range covered by the FORS2 spectroscopy (Sec. \ref{spec_class}). The complete list of parameters describing the model is compiled in Table \ref{tab:lcmodel}. We report, for each parameter and fit attempted, the median values and uncertainties to the $1\sigma$ (68\%) level derived from the marginalised posteriors (i.e 16, 50 and 84\% quantiles). If the distribution of a parameter hits against the hard limits set by a prior, we report instead an upper (lower) limit from the distribution at the 84\% (16\%) c.l.

\begin{figure}
\begin{centering}
\includegraphics[keepaspectratio, trim=7cm 0cm 3cm 13cm, clip=true, width=0.5\textwidth]{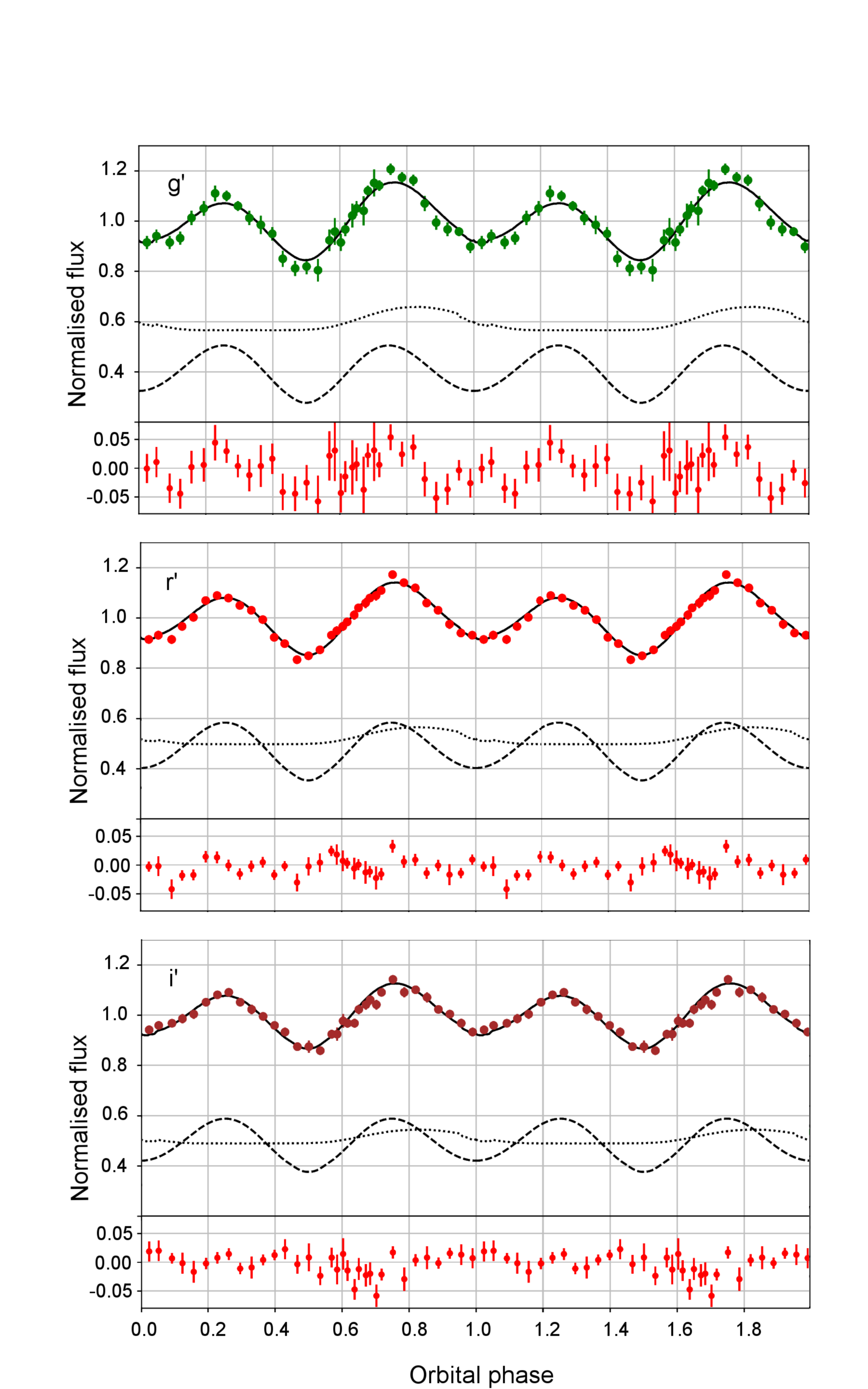}
\caption{Light curves obtained in 2014. They have been normalised in flux and phase-folded with the best orbital period and $T_0$ (two orbits are shown for clarity). The best fit for model A1 (described in Sec. \ref{photometrylc}) is plotted as a solid black line. The separate contribution to the model from the accretion disc (dotted line) and the ellipsoidal modulation (dashed line) are shown. The data is depicted as filled circles, where each panel label indicates which band they correspond to. The bottom panels are the residuals from the subtraction of the model from the observations.}
\label{fig:normphot}
\end{centering}
\end{figure}

\subsubsection{Model A1: normalised light curves}

The absolute calibration of our photometry relies on the zero-point derived for observations of a different field covered by SDSS Data Release 8 \citep{Aihara:2011aa}. To avoid potential systematic effects on the modelling due to uncertainties in the calibration, in our initial model we fitted the normalised-flux light curves of J1305, and restricted ourselves to only the high signal-to-noise data (g$^\prime$, r$^\prime$ and i$^\prime$ photometry). 

The best fit parameters and the fit to the light curve from this initial attempt (hereafter model A1) are given in Table \ref{tab:lcmodel} and shown in Fig. \ref{fig:normphot}, respectively. The model fitting statistics ($\chi^2=158/92$) show a good fit to the data. Both the companion star effective temperature $T_2=5100\pm 300\, {\rm K}$ and the photometric orbital period $P_{\rm phot}= 0.395\pm  0.002{\rm d}$ are consistent with the values measured from the spectroscopic analysis. The best fit inclination ($i=77\pm 3\, {\rm deg}$) suggests a highly inclined system in line with the detection of X-ray dips during outburst. The preferred r$^\prime$ band veiling factor from the best fit is $\sim 0.522$, about $3.5\sigma$\,lower than the value obtained from spectroscopy. The disc outer radius and temperature are within the expected range. The bright spot reproducing the asymmetric features in the light curve has a temperature of $T_{\rm spot}=6700^{+1000}_{-700}\, {\rm K}$ and a large width of at least $\zeta_{\rm width}>50\,{\rm deg}$. Inspection of the $q$ posterior distribution shows that the fit is pushing against the smallest allowed value for the mass ratio from the prior ($q>0.01$).

With the orbital inclination being close to edge-on and the assumption of a Roche Lobe filling companion, the main parameter controlling the amplitude produced by the ellipsoidal modulation is $q$, with smaller values corresponding to larger amplitudes of the light curve. We fixed the mass ratio to a reasonable value of $q=0.058$ to analyse the impact on the fitting results. As expected, the best fit under these conditions is worse than in the previous case ($\chi^2=174$). The preferred veiling factor in the r$^\prime$ band ($\sim 0.485$) is even lower than in the previous test ($\sim 0.522$).


\begin{table}
\begin{center}
\caption{Fixed parameters of the photometric modelling of J1305 common to all models. All size-related parameters are given in function of $a$.}
\label{tab:lcmodelfixed}
\begin{tabular}{lcc}

\hline
Parameter & Value & Units \\
$K_2$ & 554 & $\rm km\, s^{-1}$\\
$H_{\rm d}$ & 0.0110 & $a$\\
$r_{\rm in}$ & 0.02 & $a$\\
$n$ & 1.1 & -\\

\hline
\end{tabular}
\end{center}
\raggedright
\end{table}

\begin{table*}
\begin{center}
\caption{Parameters of the photometric modelling of J1305. All  binary size related parameters are given in function of $a$. We report for each model and parameter the median and $1\sigma$ uncertainties derived from their marginalised posteriors. If the distribution hits a prior limit, the $84$\% c.l. (or $16$\% ) is reported instead.}
\label{tab:lcmodel}
\begin{tabular}{lllll}
Free and derived parameters\\
\hline
Parameter & Model A1 & Model A2 &  Model B1 & Model B2 \\
\hline
$P_{\rm phot} ({\rm d})$ & $ 0.395\pm  0.002$ & $ 0.3958\pm  0.018$ & $0.3960 \pm 0.0017$ &  $0.395 \pm 0.002$\\
$\phi_0$ & $ 0.013 \pm 0.003$ & $ 0.013 \pm 0.002$ & $ 0.014 \pm 0.002$ & $ 0.014 \pm 0.002$\\
$i ({\rm deg})$ & $77 \pm 3$ & $72^{+5}_{-8}$& $65\pm 2$ & $56\pm 2$\\
$q$ & $<0.031$ & $0.045^{+0.022}_{-0.016}$ & $<0.032$ & $0.056^{+0.024}_{-0.019}$ \\
$T_{2} \rm{(K)}$ & $ 5100\pm 300$ &  $ 6500^{+900}_{-1100}$ &  $ 4250\pm 100$ & $ 4700^{+200}_{-300}$\\
$T_{\rm edge} \rm{(K)}$ & $5500\pm 500$ & $ 6800^{+1400}_{-1200}$ & $4500\pm 170$ & $4900\pm 500$\\
$R_{\rm d} (a)$ & $0.4\pm 0.1$ & $ < 0.4$ & $>0.5 $ &  $ < 0.4$ \\
$T_{\rm spot} \rm{(K)}$ & $6700^{+1000}_{-700}$ & $ 9100^{+1100}_{-1800}$ & $4900\pm 200$ &  $6400\pm 800$ \\
$\zeta_{\rm mid} ({\rm deg})$ & $113\pm 5$  & $117\pm 5$ & $115^{+8}_{-4}$ & $121\pm 4$  \\
$\zeta_{\rm width} ({\rm deg})$ & $>50$ & $>70$ & $>70$ &$>60$ \\
$\rm E(B-V) (mag)$ & $-$ & $-$ & $0.14 \pm 0.04$ & $0.34 \pm 0.10$\\

\hline
$L_{\rm d}$ (erg/s) & $1.4^{+1.0}_{-0.6} \cdot  10^{33} $ & $< 4 \cdot  10^{32}$ & $< 9 \cdot  10^{31} $ &  $1.4\pm 0.7 \cdot  10^{32}$ \\
$M_1\, \rm{(M_\odot)}$ & $8.0\pm 0.2$  & $9.1^{+1.5}_{-1.0}$ & $10.0\pm 0.4$ &  $14.1^{+0.8}_{-1.0}$ \\
$M_2\, \rm{(M_\odot)}$ & $0.22^{+0.05}_{-0.02}$  & $0.43^{+0.31}_{-0.16}$ & $0.29\pm 0.04$  & $0.9\pm 0.4$ \\
$\chi^2/{\rm d.o.f}$ & $158/92$ & $126/92$ &  $161/94$ & $127/94$ \\
\hline
\end{tabular}
\end{center}
\raggedright
\end{table*}




\subsubsection{Model B1: flux-calibrated light curves}

In a second attempt, we fit the flux-calibrated photometry, where the GROND efficiencies were used to transform magnitudes in to flux densities.
We also allowed for different reddening factors in each band, using the linear transformation of extinction coefficients between bands for GROND filters ($A_{g^{'}}/A_V=1.255$, $A_{r^{'}}/A_V=0.866$, $A_{i^{'}}/A_V=0.648$; similar to those reported for SDSS primed filters in \citealt{Schlafly2011}) to define the correction with a single parameter: $A_V=3.08\, {\rm E(B-V)}$. 

Our best fit results from this model (hereafter model B1) are presented in Tab. \ref{tab:lcmodel}. The orbital period $P_{\rm phot}= 0.3960\pm  0.0017{\rm d}$ is consistent with both the spectroscopic and the photometric period from model A1. The effective temperature of the companion star is cooler than that preferred in model A1, with the spectroscopic value falling in between these two results. The hydrogen column density derived from the non-dip X-ray spectrum of the source ($N_{\rm H}= 1.0\cdot 10^{21}\, {\rm cm^{-3}}$, \citealt{Shidatsu2013}) can be employed to derive a reddening of $\rm{E(B-V)}\sim 0.18$\,mag \citep{Predehl1995}. This is fully consistent with the preferred value obtained in our light curve modelling.
The accretion disc edge temperature appears colder than that obtained from model A1, but the disc radius distribution is found to ram against the hard upper limit imposed by the prior. With regard to the bright spot, a feature with a large width spread along the edge of the disc is still preferred, with a temperature just slightly higher than the edge temperature. The model favours again a mass ratio as small as allowed by the priors, showing that model B1 was not able to constrain this parameter either. The preferred orbital inclination ($i=65\pm 2\, {\rm deg}$) is lower than that of the previous model, while the disc luminosity is favouring values as small as possible. The r$^\prime$ band veiling factor for the best fit is $\sim 0.497$, in line with model A1 in favouring lower values than the spectroscopic measurement.

In order to investigate the effect of $q$ on our results, we repeated the fit but fixing the mass ratio to two extreme values: $q=0.025$ and $q=0.07$. As expected, the higher $q$ value corresponds to a worse fit ($\chi^{2}=179$), but most of the parameters remain the same as for model B1. Only the disc luminosity and radius ($L_{\rm d}\sim 1.6\cdot 10^{32}\, {\rm erg s^{-1}}$, $R_d\sim 0.43\, a$) appear better constrained.

\subsubsection{Model A2 and B2: relaxing the prior on the disc veiling}

The models inspected above showed consistent results for some of the best fit parameters, but the mass ratio in particular remained elusive. Adding other photometric epochs to the fit did not improve the results due to both the scarce phase coverage and the worse seeing. However, we noticed that the veiling factor in models A1 and B1 always prefers significantly lower values than the spectroscopic one. For this reason, we decided to attempt a new fit relaxing the constraints on the veiling factor for the r$^\prime$ band. The fact that the photometric and spectroscopic datasets were obtained two years apart, combined with a variable contribution of the accretion disc within yearly timescales (a known effect seen in other quiescent LMXBs, see e.g. \citealt{Cantrell2008}) is enough to warrant this situation.

We refer hereafter as model A2 and B2 to models analogous to A1 and B1, respectively, but removing the veiling prior during the fit. An immediate, common result to both fits is the much better statistics ($\chi^2/{\rm d.o.f}$=126/92 and 127/94) compared with the original version of the models, which shows to what extent the prior on the veiling factor affected our results. Model A2 best fit parameters are close to those derived for model A1. The obtained inclination is slightly lower ($72^{+5}_{-8}\, \rm{deg}$) while the companion star $T_2$ has increased, though the larger uncertainties make these results still consistent within $1\sigma$ with the former. One of the key differences between models A1 and A2 is that the later has a posterior distribution for $q$ that is not pushing against the lower limits from the prior. While the preferred value $q=0.045^{+0.022}_{-0.016}$ is not particularly constraining, it perfectly fits within the expected range. The second key difference between these models are all the parameters associated with the accretion disc, which varied to account for the different veiling factor. The veiling preferred by model A2 in r$^\prime$ band is $X=0.28\pm0.07$. 

On the other hand, model B2 fit prefers an even lower inclination than the original flux-calibrated model (B1). The reddening parameter is higher than that predicted by model B1, therefore inconsistent with the value derived from X-ray spectroscopy. It shows a similar mass ratio to that preferred by model A2, suggesting that in the previous models (A1 and B1) the determination of $q$ was largely hampered by the veiling factor prior. On this topic, both model A2 and B2 best fits give a similarly low veiling factor ($\sim 0.3$). We discuss the disagreement between the spectroscopic and photometric veiling factors in Sec. \ref{discveil}. 

After analysis of the models presented in this section, we will hereafter consider the photometric results from model A2 best fit. The better $\chi^2$ of the model, as well as the inclination uncertainty making it consistent with both models A1 and B1, led us to select it as the best fit to the data. We note that these results must be treated with caution, as they might be affected by incompleteness of the tested models, as well as by the limited signal-to-noise ratio of our observations.

\subsection{Constraints on the disc veiling factor}
\label{discveil}

We have obtained different veiling factors from the spectroscopic classification ($X_{\rm spec}=0.66\pm 0.04$) and from the best photometric fit (model A2, $X_{\rm phot}=0.28\pm 0.07$). To understand this, it is worth noticing the stationary absorption features found in the spectra (peak $A$ from the CCFs), which we associate with contamination from a nearby star. On this note, we discussed the potential influence of the nearest resolved field star in Sec. \ref{observations}, and concluded it should not contribute significantly to the total flux observed if the slit is properly centred on the target. However, a drift of the slit position towards the North of J1305 during the observations would increase the relative contribution of this field star, and potentially be the cause of peak $A$. If we consider the slit is always perfectly centred, then the origin should be a second interloper in the line of sight, unresolved even by Gaia (spatial resolution of $\sim 0.4''$, \citealt{GaiaDR22018}). To investigate this, we perform a spectral classification of the averaged spectrum in the observer rest frame (following the same steps as in Sec. \ref{spec_class}), and obtain $T_{\rm eff}=4240 \pm 80 {\rm K}$, consistent with a dwarf star interloper of spectral type K6-7 \citep{Pecaut2013}. More interestingly, the fractional flux contribution of this interloper to the total flux in the averaged spectrum is $0.27\pm 0.02$. This effectively implies that the veiling to J1305 flux associated with the accretion disc after excluding the interloper contribution during 2016 epoch should be $X_{\rm spec}= 0.53\pm 0.06$. Such a disc veiling fraction is easier to reconcile with the best fit to 2014 photometric observations ($X_{\rm phot}=0.28\pm 0.07$). If the remaining difference between the veiling factor of 2014 and 2016 epochs is due to intrinsic variability of the accretion disc, it would appear as a slight brightening of the system of $0.49\pm 0.18 \,\rm{mag}$. This result is just about consistent with the $ 0.14 \pm 0.09\,\rm{mag}$ brightening observed (Tab. \ref{tab:GROND_lateLog}) at $1.5\sigma $. On the contrary, assuming the original veiling factor from the spectroscopy is intrinsic to the system ($0.66\pm 0.04$) requires a larger brightening ($0.82\pm 0.17\,\rm{mag}$), which seems unlikely given the photometric observations. This argument supports the nearby star at $1.64''$ as the origin of the static absorption lines producing peak $A$: while the 2014 photometric observations perfectly resolve this interloper from the target, a shift in the slit centre during 2016 spectroscopy could led to the observed contamination. To further investigate this hypothesis, we extracted the apparent magnitudes of this nearby star during the 2014 epoch, and found g$^\prime=22.83\pm 0.03$, r$^\prime=21.27\pm 0.01$, i$^\prime=20.46\pm 0.01$ and z$^\prime=19.97\pm 0.01$. Adopting a similar distance to J1305, we then correct the above magnitudes using the same reddening value considered for J1305. The obtained colours suggest a spectral type for the nearby star of $\sim$M0 V \citep{Pecaut2013}, slightly later than that derived for the spectroscopic interloper, but still consistent if the reddening (distance) is lower. The continuum contribution from this interloper to the observed spectra would also decrease the measured equivalent width of the $\rm H\alpha$ line (${\rm EW}=-21\pm 3\,$\AA). Indeed, the measured value in J1305 is lower than those observed in other high inclination LMXBs (e.g., Swift J1357.2-0933, \citealt{Torres2015,MataSanchez2015b};  MAXI J1820+070, \citealt{Torres2019,Torres2020}; MAXI J1659-152 \citealt{Torres2021}), further favouring this scenario.

\section{Discussion}
\label{discussion}

The CCFs retrieved by comparing the observed spectra with late-type template stars are double-peaked, associated with two sets of absorption features similar to that of the template star. While peak $A$ was identified as contamination from an interloper star, the temporal evolution of peak $B$ was associated with the motion of the companion star, leading to $K_2$ and $P_{\rm spec}$. We will employ hereafter the spectroscopic period as the orbital period of the binary ($P_{\rm orb}=0.394\pm 0.004\, {\rm d}$), as this is consistent with the $P_{\rm phot}$ derived in Sec. \ref{photometrylc}. The closest dipping period proposed in the literature ($P_{\rm dip}=0.4058\pm 0.0017\,{\rm d}$, \citealt{Shidatsu2013}) is slightly higher than the $P_{\rm orb}$ found in this work. However, the former dipping periodicity was obtained by using the temporal separation between three "deep dips" detected in their light curve rather than performing a traditional timing analysis. A detailed study by \citet{Shaw2017} showed multiple possible periodicities as a result of the complexity of the X-ray light curve. While their favoured dipping period ($0.208\pm 0.002\, {\rm d}$) is even further away from the $P_{\rm orb}$ presented here, it shows the risks of associating periodicities detected during the outburst with the true $P_{\rm orb}$ of the system.

The determination of $K_2$ and $P_{\rm orb}$ allow us to impose a lower limit to the compact object mass via the mass function:

$$f(M_1)=\frac{M_1\, \sin i^3}{(1+q)^2}=\frac{P_{\rm orb}K_2^3}{2\pi G}=6.9\pm 0.3\, M_{\odot}$$

The most conservative estimate for the compact object mass is comfortably above the predicted maximum mass for a neutron star ($\lesssim 3\, M_{\odot}$, e.g., \citealt{Kalogera1996}), allowing us to dynamically confirm the BH nature of J1305 for the first time. To solve the dynamics of the system, the missing key parameters are the mass ratio, which is not well constrained from the photometric modelling, and the orbital inclination.

From our spectroscopy, we obtained $v \sin i <110\, {\rm km\, s^{-1}}$ (c.l. $84\%$), which implies an upper limit to the mass ratio of $q<0.07$. We also apply a conservative lower limit of $q>0.01$, the same prior we applied for the photometric modelling and consistent with even the most extreme mass ratio measured in an LMXB (XTE J1118+480, $q=0.024\pm 0.009$, \citealt{Calvelo2009}). On the other hand, the best photometric model (A2) favours $q=0.05\pm 0.02$, which is not particularly constraining but it is consistent with the aforementioned limits. Given that $q=K_1/K_2$, another possibility for deriving $q$ relies on measuring $K_1$. We employed the diagnostic diagram method (see \citealt{Shafter1986}), which tracks the movement of the wings of the emission line profiles formed in the disc (arising from the innermost regions of the accretion disc) as a proxy for the compact object movement. Unfortunately, our results do not reveal any clear periodic evolution of the $\rm H\alpha$ line, hampering our efforts to obtain an estimation of $q$. \citet{Casares2016} presented a correlation between two parameters derived from the $\rm H\alpha$ line profile ($\rm FWHM$ and the peak to peak separation, ${\rm DP}$) and the mass ratio of the X-ray binary. Both ${\rm DP}$ and $\rm FWHM$ are measured directly on the averaged spectrum without applying any radial velocity shift to the individual spectra. Masking out the core of the line (to avoid potential contamination by the interloper), we find ${\rm FWHM}=2368 \pm  13 \, {\rm km\, s^{-1}}$ and ${\rm DP}=1378\pm 4 \, {\rm km\, s^{-1}}$. Following the equation presented in \citet{Casares2016}, we derive $q\sim 0.038$ (corresponding to $v\sin{i}\sim 88 \, {\rm km\, s^{-1}}$ and $K_1\sim 21\, {\rm km\, s^{-1}}$), well within the range defined by our photometric fit. We will adopt hereafter a normal distribution for the mass ratio of $q=0.05\pm 0.02$ but truncated to $0.01<q<0.07$. The masses for the components of the binary can be then set in terms of the orbital inclination:

$$M_1=\dfrac{7.6\pm 0.4}{\sin{i}^3}\, M_\odot; \qquad M_2=\dfrac{0.35\pm 0.12}{\sin{i}^3}\, M_\odot$$

The orbital inclination has the strongest effect on the derived BH mass. Our best light curve models A2 and B2 yield barely consistent values of $72^{+5}_{-8}\, {\rm deg}$ and $56\pm 2\, {\rm deg}$, respectively. The detection of X-ray dips in J1305 also allows us to constrain its inclination to be at least $i>60\, {\rm deg}$ \citep{Frank1987}. Geometrical arguments can be employed to put an upper limit to the inclination, as eclipses by the companion star of the central X-ray source are not observed and the X-ray spectrum shows a thermal component (i.e. it is not an accretion disc corona source). This implies: $\cos{i}\geq R_2/a$. As the companion is assumed to be Roche Lobe filling for the transfer of mass to occur, we use the volume averaged Roche Lobe radius \citep{Eggleton1983}: $ R_2/a\approx R_{L2}/a\sim 0.16$, which implies $i<82\, {\rm deg}$ (c.l. 84\%). There is no clear evidence for eclipses in the optical curve either, including grazing eclipses similar to those observed in MAXI J1820+070 \citep{Torres2019}. The constraints on the inclination ($i=60-82\, {\rm deg}$) only allow us to rule out the photometric model B2, while the remaining (A1, A2 and B1) each favours a different extreme of the range.

 Absorption features not related with the companion star have previously been observed in cataclysmic variables (CVs), binaries formed by a white dwarf and a companion star. In particular, they manifest as deep, narrow cores embedded in broader disc emission lines (mainly the Hydrogen Balmer series), with radial velocity curves inconsistent with neither the companion star or the white dwarf itself. They have been observed in high inclination cataclysmic variables ($ i\gtrsim 75\, {\rm \deg}$; \citealt{Schoembs1983}), and they are associated with the occultation of the inner accretion disc by structures in the outer rim (e.g., dwarf nova Z Chamaeleontis, see \citealt{Marsh1987}). High inclination LMXBs also show deep and relatively narrow cores during quiescence, such as Swift J1357.2-0933 \citep{Torres2015,MataSanchez2015b} and MAXI J1659-152 \citep{Torres2021}. There are no reports of the radial velocity curves from these cores producing a clear periodic signature. Nevertheless, double peak profiles of accretion disc emission lines are already expected to become broader and their core deeper for higher inclinations, as the peaks become more separated \citep{Horne1986}. In both of these cases, deep cores embedded into the emission lines are considered a clear indicator of a high orbital inclination. We notice a deep and narrow absorption core overlapping the profile depression delimited by the two peaks of the $\rm H\alpha$ line in the observer rest frame spectrum of J1305, reaching the continuum level and below. The evolution during the orbit of this core shows the deepest examples (reaching down to 60\% of the continuum) at superior conjunction (phase $0.5$, with the BH between the companion star and the Earth), while at inferior conjunction they are shallower. The companion star cannot be the origin of this narrow feature either, due to its late spectral type and the expected smearing in the observer rest frame. The interloper star could produce a similar apparent feature, but its late spectral type and small fractional contribution to the total light are not enough to explain the depths of the observed core. On this basis, we will favour hereafter the orbital inclination from model A2 ($72^{+5}_{-8}\, {\rm deg}$). This implies BH and companion masses of: $M_1= 9.0^{+1.5}_{-1.0}\, M_{\odot}$ and $M_2= 0.43\pm 0.16\, M_{\odot}$. Nevertheless, to account for the mixed results from the photometric modelling, we also present the resulting mass for the conservative constraint on the inclination from the X-ray dips: $M_1= 8.9_{-1.0}^{+1.6}\, M_{\odot}$ and $M_2= 0.43\pm 0.16\, M_{\odot}$. The dynamical BH mass is remarkably higher than the $M_1\lesssim 4\, M_{\odot}$ derived from previous X-ray spectral modelling (\citealt{Morihana2013}, using $i<82\, {\rm deg}$ and a distance of $d\sim 10\,{\rm kpc}$). Their model assumes a non-spinning BH and an accretion disc inner radius reaching the innermost stable orbit (ISCO). In order to reconcile both measurements, the BH should have a retrograde spin, effectively increasing the ISCO radius. Alternatively, a revisit of the X-ray spectrum employing more complex models (e.g. including electron scattering effects in the disc atmosphere) might change the measured inner radius of the accretion disc, potentially easing the tension with the results presented here, though such a study might be hampered by the limited signal-to-noise of the X-ray data.

\subsection{The companion star of J1305}
\label{disc:companion}

Under the assumption that the donor star fills its Roche Lobe and is tidally locked, a correlation between the orbital period and the mean density exists \citep{Faulkner1972}. For $P_{\rm orb}=0.394\pm 0.004\, {\rm d}$, this implies an early G spectral type for the companion if it is a dwarf on the main sequence. On the other side, the spectral classification of the companion (based on the comparison with dwarf star templates) favours a spectral type K3-5 ($T_{\rm eff}=4610^{+130}_{-160}\, {\rm K}$, see Sec. \ref{spec_class}), which for a dwarf companion would imply a mass of $0.74 \pm 0.04  \, M_{\odot}$ and a stellar radius of $0.73 \pm 0.03\, {R_\odot} $ \citep{Pecaut2013}. Additionally, the previous dynamical analysis, which was based on our best determinations for the mass ratio and the inclination, results in a stellar radius of $R_2= 0.73 \pm 0.09\, {R_\odot}$ \citep{Eggleton1983}, similar to the expected value for K3-5 type, but a much lower mass of only $M_2= 0.43\pm 0.16\, M_{\odot}$. The discrepancy between the predicted spectral type and mass of the companion has been previously reported for several LMXBs, being an indication that the companion is slightly evolved compared to a main sequence member (see \citealt{Rappaport1983}; \citealt{Podsiadlowski2003}). Thus, the mass proposed from the spectral class should be considered as an upper limit to the true value. Theoretical studies of the evolution of the companion star in compact binaries using stripped giant models have shown that its properties mainly depend on the mass of its Helium core. On this topic, \citet{King1993} presented a formula that allows to put constraints on the mass of a tidally locked evolved companion from its $P_{\rm orb}$, which for J1305 results in $0.14\, M_\odot <M_2<0.9 \, M_\odot$. The formula also predicts a range of effective temperatures ($5000-5400\, {\rm K}$) and radius ($0.5<R_2<1.0 \, R_\odot$) for the companion. While the mass range is consistent with the value previously proposed, the predicted range in effective temperature is slightly higher than our spectral classification suggests. The upper limit on $T_{\rm eff}$ (i.e. minimum $M_2$ and $R_2$) is quite conservative, as it assumes the extreme case of the companion being a naked Helium core. On the other hand, the lower limit on $T_{\rm eff}$ (i.e. maximum $M_2$ and $R_2$) is derived from the Sch\"onberg-Chandrasekhar limit, a critical mass ratio between the core and the total mass of the companion ($M_c/M_2<q_{\rm crit}$) above which an isothermal core collapses \citep{SC1942}. The formula presented in \citet{King1993} holds for a critical value of $q_{\rm crit}=0.17$, but other authors have proposed different limits (e.g., $q_{\rm crit}=0.10$, \citealt{Beech1988}, \citealt{Eggleton1998}). The most recent studies using state-of-the-art evolutionary models (instead of analytical, polytropic models) have suggested that this limit might not be a sharp cut-off but rather a smooth transition (e.g., \citealt{Ziolkowski2020}). For all these reasons, the lower limit on $T_{\rm eff}$ (hence the upper limit on $M_2$) here presented must be taken with caution rather than employing it as a strict constraint. In addition, it is worth remarking here that the spectral classification in this paper was performed against synthetic dwarf star templates which are not expected to perfectly reproduce the J1305 evolved companion. All things considered, we put conservative constrains on the mass of the companion star of $0.14\, M_\odot <M_2<0.78 \, M_\odot$. This conservative range is consistent with the derived value from the dynamical analysis ($M_2= 0.43\pm 0.16\, M_{\odot}$), and confirms the evolved nature from the donor star as well as support its stripped giant nature as a consequence of mass transfer on to the BH.

\subsection{Distance and proper motion}
\label{distance}

An optical counterpart of J1305 has been detected by the Gaia mission \citep{GaiaMission2016} but precise determination of its parallax was not possible, partly due to its optical faintness. One can adopt a dwarf star companion as a first approach to derive the distance to the system. For this purpose, we will consider the mean observed r$^\prime$-band magnitude of J1305 during the 2016 photometric epoch ($20.48\pm 0.07$), which is simultaneous with our spectroscopy. We correct this from the extinction in this band ($A_{r'}=0.37\pm 0.11$), as well as from the effect of the disc veiling ($\Delta r'_{veil}=-2.5 \log{(1-X)}$; we use $X=0.53 \pm 0.06$, see Sec. \ref{discveil}) to obtain an apparent magnitude for the companion star of $r'=20.96\pm 0.20$. Direct comparison with tabulated absolute magnitudes for dwarf stars of the expected spectral type (K3-5, \citealt{Pecaut2013}) after transformation to the r$^\prime$ band \citep{Jordi2006} results in a distance to the system of $7.5^{+1.8}_{-1.4}\, {\rm kpc}$, where we have considered the uncertainties on the spectral type, the veiling factor, the extinction and the observed magnitude. If we employ instead for the comparison a black-body spectral energy distribution for the measured $T_{\rm eff}\sim 4610\, {\rm K}$ and the companion radius previously derived $R_2\sim 0.7\, R_\odot$, we calculate the emitted flux density at the star surface. We do this for the r$^\prime$ band by integrating the product of the emitted flux density by the corresponding GROND filter transmission profile over the wavelength dimension (Spanish Virtual Observatory\footnote{\url{http://svo2.cab.inta-csic.es/theory/fps/}}, SVO), and normalising it by the transmission profile integral. The resulting $f_{\nu}(r')\sim 13\,{\rm \mu  Jy} $ implies a fully consistent distance with the former of $\sim 8\, {\rm kpc}$.

Our analysis results in a distance to J1305 of $7.5^{+1.8}_{-1.4}\, {\rm kpc}$, conditioned to the assumptions previously described. Previous attempts on determining the distance to J1305 from spectral X-ray modelling \citep{Shidatsu2013} proposed $d=6.3^{+0.4}_{-0.3}\, {\rm kpc}$. They assumed both a BH mass of $3\, M_\odot$ and a bolometric luminosity in terms of the Eddington luminosity of $ 0.05\, L_{\rm edd}$. While reproducing their X-ray modelling is beyond the scope of this paper, the increase of $L_{\rm edd}$ using the BH mass derived in this work already suggests a larger distance, easier to reconcile with our results.

Combination of the distance with the Galactic latitude of the system results in a height under the Galactic plane of $z=-1.0\pm 0.2\, {\rm kpc}$. This potentially places J1305 within the Galactic thick disc ($|z|\gtrsim 1\, {\rm kpc}$, \citealt{Gilmore1983}), while most of BH LMXBs are typically found much closer to the Galactic plane (see e.g., \citealt{Corral-Santana2016}). With the determination of the distance towards J1305, we are in a position to calculate its space velocity. The latest Gaia \citep{GaiaMission2016} Early Data Release 3 astrometric solution for the optical counterpart \citep{GaiaEDR32020,GaiaEDR32020b} provides a proper motion mainly aligned in the direction of the right ascension coordinate: $\Delta \alpha = -7.89 \pm 0.62\,{\rm mas\, yr^{-1}}$ and $\Delta \delta=-0.16 \pm 0.72\,{\rm mas \, yr^{-1}}$. At the distance and systemic velocity derived ($\gamma=-9\pm 5\, {\rm km\, s^{-1}}$, Sec. \ref{rvcurve}), the space velocity in the local standard of rest for J1305 results $(U,V,W)=(-220\pm 50,-140\pm 40, 20 \pm 30)\, {\rm km \, s^{-1}}$, using the transformations given by \citet{Johnson1987}, and assuming a Solar space velocity in this frame of $(U_\odot,V_\odot,W_\odot)=(9,12, 7)\, {\rm km \, s^{-1}}$ \citep{Mihalas1981}. We compare this space velocity with that determined for stars at the same Galactocentric coordinates but projected onto the Galactic plane. For this, we use the galactic dynamics package \textsc{galpy} \citep{Bovy2015} with the Milky Way potential \textsc{MWPotential2014} to obtain a peculiar velocity for J1305 of $80\pm 30 \, {\rm km \, s^{-1}}$. From this result, J1305 is moving away from the Galactic centre, rotating about it with a slower velocity than the stars in the projected vicinity onto the Galactic disc, and without a significant component perpendicular to the Galactic plane. We note that this comparison must be taken with caution due to the $z=1.0\pm 0.2\, {\rm kpc}$ of J1305. Previous studies of the thin/thick disc population (e.g. \citealt{Chiba2000}, \citealt{Soubiran2003}) suggest they have lower rotational velocities than stars in the Galactic plane. Similarly high spatial velocities to that found for J1305 ($v_{\rm space}=270\pm 70 \, {\rm km\, s^{-1}}$) have been observed for other LMXBs harbouring neutron stars (e.g. Cen X-4, \citealt{GonzalezHernandez2005}), and they have been associated with strong natal kicks during the compact object formation by tracing back their Galactocentric orbits. This scenario is not that clear for BHs, where both low-velocity (e.g. GRS 1915+105, \citealt{Dhawan2007}) and high velocity systems (e.g. XTE J1118+480, \citealt{Mirabel2001}; GRO J1655-40, \citealt{Willems2005}) have been discovered. The application of the code described in \citet{Atri2019} to J1305 allows us to integrate its Galactocentric orbit back in time and calculate a distribution of potential natal kick velocities. We obtain a median value of $v_{\rm kick}=75^{+21}_{-12}\, {\rm km\, s^{-1}}$ (68\%), suggesting a significant natal kick for the system consistent with the known population (see \citealt{Atri2019} and references therein).

\section{Conclusions}
\label{conclusion}

We have presented the first optical study of MAXI J1305-704 in quiescence using both photometric and spectroscopic observations. Analysis of both observational sets allows us to measure an orbital period of $P_{\rm orb}= 0.394\pm  0.004\,{\rm d}$, barely consistent with the X-ray dipping period proposed by \citet{Shidatsu2013}, and ruling out all other proposed values from X-ray studies. This is a cautionary tale on the association of periodic X-ray variability with the orbital period, and remarks the decisive role of optical observations on this debate. The CCFs of the observed spectra with dwarf star templates yields a radial velocity semi-amplitude for the companion star of $K_2=554\pm 8\, {\rm km\, s^{-1}}$, as well as a radial systemic velocity of $\gamma =-9\pm 5\, {\rm km\, s^{-1}} $ and a spectral type for the companion star of K3-5. In addition, we detect another set of absorption features apparently stationary, which we propose are due to contamination from an interloper star. 

The compact object mass function of $f(M_1)=6.9\pm 0.3\, M_{\odot}$ allows us to confirm, for the first time, its BH nature. From modelling of the optical light curve, we favour an orbital inclination of $72^{+5}_{-8}\, {\rm deg}$ and a mass ratio of $q=0.05\pm 0.02$. This is consistent with the high inclination scenario previously proposed from X-ray dipping phenomenology, which establishes a conservative range of $60\, {\rm deg}<i<82\, {\rm deg}$. The latter range results in a BH mass of $M_1= 8.9_{-1.0}^{+1.6}\, M_{\odot}$ and a companion mass of $M_2= 0.43\pm 0.16\, M_{\odot}$. Under the assumption of a dwarf companion, we obtain a distance for J1305 of $d=7.5^{+1.8}_{-1.4}\, {\rm kpc}$. When combined with the Galactic coordinates of the source and its proper motion, we find J1305 is probably a member of the Galactic thick disc, and shows a high spatial velocity ($v_{\rm space}=270\pm 70 \, {\rm km\, s^{-1}}$) tentatively associated with a significant natal kick.

\section*{Acknowledgements}
\label{acknowledgements}

 We are grateful to James C. Miller-Jones and Tyrone O'Doherty for insightful conversations about the natal kick velocity. We also thank the anonymous referee for the constructive remarks. DMS acknowledges support from the ERC under the European Union's Horizon 2020 research and innovation programme (grant agreement no. 715051; Spiders). This work has been supported in part by the Spanish MINECO under grant AYA2017-83216-P. MAPT acknowledges support via a Ram\'on y Cajal Fellowship RYC-2015-17854. Part of the funding for GROND (both hardware as well as personnel) was generously granted from the Leibniz-Prize to Prof. G. Hasinger (DFG grant HA 1850/28-1). Based on observations collected at the European Southern Observatory under ESO programmes 093.A-9099(A), 096.D-0430(A), 096.A-9025(A), and 0101.A-9099(A). We are grateful to T.-W. Chen, A. Hempel, M. Millon, and H. Steinle for performing the GROND observations.  
This work has made use of data from the European Space Agency (ESA) mission {\it Gaia} (\url{https://www.cosmos.esa.int/gaia}), processed by the {\it Gaia} Data Processing and Analysis Consortium (DPAC, \url{https://www.cosmos.esa.int/web/gaia/dpac/consortium}). Funding for the DPAC has been provided by national institutions, in particular the institutions participating in the {\it Gaia} Multilateral Agreement. This research has made use of the SVO Filter Profile Service supported from the Spanish MINECO through grant AYA2017-84089.

\section*{Data availability}
\label{dataavailability}

The data underlying this article are publicly available at \url{http://archive.eso.org/cms.html}. Spectroscopic observations can be found under program ID 096.D-0430(A). The photometric data are also publicly available at the same location under program IDs 092.A-9099(A) and 096.A-9025(A). The remaining data corresponding to the 2018 photometric observations were obtained under a different program ID 0101.A-9099(A), and will be made public after the release date in June 2028.

\bibliographystyle{mnras} 
\bibliography{bibliography.bib}


\bsp	
\label{lastpage}
\end{document}

%% file: def_bibtex.tex
\def\apj{ApJ}
\def\mnras{MNRAS}
\def\aap{A\&A}
\def\apjl{ApJ}
\def\pasj{PASJ}
\def\nat{Nature}

\def\apss{Ap\&SS}
\def\apjs{ApJS}

\def\pasp{PASP}
\def\aj{AJ}
\def\ssr{SSR}

%% file: MAXIJ1305.bbl
\begin{thebibliography}{}
\makeatletter
\relax
\def\mn@urlcharsother{\let\do\@makeother \do\$\do\&\do\#\do\^\do\_\do\%\do\~}
\def\mn@doi{\begingroup\mn@urlcharsother \@ifnextchar [ {\mn@doi@}
  {\mn@doi@[]}}
\def\mn@doi@[#1]#2{\def\@tempa{#1}\ifx\@tempa\@empty \href
  {http://dx.doi.org/#2} {doi:#2}\else \href {http://dx.doi.org/#2} {#1}\fi
  \endgroup}
\def\mn@eprint#1#2{\mn@eprint@#1:#2::\@nil}
\def\mn@eprint@arXiv#1{\href {http://arxiv.org/abs/#1} {{\tt arXiv:#1}}}
\def\mn@eprint@dblp#1{\href {http://dblp.uni-trier.de/rec/bibtex/#1.xml}
  {dblp:#1}}
\def\mn@eprint@#1:#2:#3:#4\@nil{\def\@tempa {#1}\def\@tempb {#2}\def\@tempc
  {#3}\ifx \@tempc \@empty \let \@tempc \@tempb \let \@tempb \@tempa \fi \ifx
  \@tempb \@empty \def\@tempb {arXiv}\fi \@ifundefined
  {mn@eprint@\@tempb}{\@tempb:\@tempc}{\expandafter \expandafter \csname
  mn@eprint@\@tempb\endcsname \expandafter{\@tempc}}}

\bibitem[\protect\citeauthoryear{{Aihara} et~al.,}{{Aihara}
  et~al.}{2011}]{Aihara:2011aa}
{Aihara} H.,  et~al., 2011, \mn@doi [\apjs] {10.1088/0067-0049/193/2/29}, \href
  {https://ui.adsabs.harvard.edu/abs/2011ApJS..193...29A} {193, 29}

\bibitem[\protect\citeauthoryear{{Al-Naimiy}}{{Al-Naimiy}}{1978}]{AlNaimiy1978}
{Al-Naimiy} H.~M.,  1978, \mn@doi [\apss] {10.1007/BF00645913}, \href
  {http://adsabs.harvard.edu/abs/1978Ap%26SS..53..181A} {53, 181}

\bibitem[\protect\citeauthoryear{{Appenzeller} et~al.,}{{Appenzeller}
  et~al.}{1998}]{Appenzeller1998}
{Appenzeller} I.,  et~al., 1998, The Messenger, \href
  {https://ui.adsabs.harvard.edu/abs/1998Msngr..94....1A} {94, 1}

\bibitem[\protect\citeauthoryear{{Archibald} et~al.,}{{Archibald}
  et~al.}{2009}]{Archibald2009}
{Archibald} A.~M.,  et~al., 2009, \mn@doi [Science] {10.1126/science.1172740},
  \href {https://ui.adsabs.harvard.edu/abs/2009Sci...324.1411A} {324, 1411}

\bibitem[\protect\citeauthoryear{{Atri} et~al.,}{{Atri}
  et~al.}{2019}]{Atri2019}
{Atri} P.,  et~al., 2019, \mn@doi [\mnras] {10.1093/mnras/stz2335}, \href
  {https://ui.adsabs.harvard.edu/abs/2019MNRAS.489.3116A} {489, 3116}

\bibitem[\protect\citeauthoryear{{Bayless}, {Robinson}, {Hynes}, {Ashcraft}  \&
  {Cornell}}{{Bayless} et~al.}{2010}]{Bayless2010}
{Bayless} A.~J.,  {Robinson} E.~L.,  {Hynes} R.~I.,  {Ashcraft} T.~A.,
  {Cornell} M.~E.,  2010, \mn@doi [\apj] {10.1088/0004-637X/709/1/251}, \href
  {https://ui.adsabs.harvard.edu/abs/2010ApJ...709..251B} {709, 251}

\bibitem[\protect\citeauthoryear{{Beech}}{{Beech}}{1988}]{Beech1988}
{Beech} M.,  1988, \mn@doi [\apss] {10.1007/BF00645666}, \href
  {https://ui.adsabs.harvard.edu/abs/1988Ap&SS.147..219B} {147, 219}

\bibitem[\protect\citeauthoryear{{Belloni}, {Motta}  \&
  {Mu{\~n}oz-Darias}}{{Belloni} et~al.}{2011}]{Belloni2011}
{Belloni} T.~M.,  {Motta} S.~E.,   {Mu{\~n}oz-Darias} T.,  2011, Bulletin of
  the Astronomical Society of India, \href
  {https://ui.adsabs.harvard.edu/abs/2011BASI...39..409B} {39, 409}

\bibitem[\protect\citeauthoryear{{Bovy}}{{Bovy}}{2015}]{Bovy2015}
{Bovy} J.,  2015, \mn@doi [\apjs] {10.1088/0067-0049/216/2/29}, \href
  {https://ui.adsabs.harvard.edu/abs/2015ApJS..216...29B} {216, 29}

\bibitem[\protect\citeauthoryear{{Calvelo}, {Vrtilek}, {Steeghs}, {Torres},
  {Neilsen}, {Filippenko}  \& {Gonz{\'a}lez Hern{\'a}ndez}}{{Calvelo}
  et~al.}{2009}]{Calvelo2009}
{Calvelo} D.~E.,  {Vrtilek} S.~D.,  {Steeghs} D.,  {Torres} M.~A.~P.,
  {Neilsen} J.,  {Filippenko} A.~V.,   {Gonz{\'a}lez Hern{\'a}ndez} J.~I.,
  2009, \mn@doi [\mnras] {10.1111/j.1365-2966.2009.15304.x}, \href
  {https://ui.adsabs.harvard.edu/abs/2009MNRAS.399..539C} {399, 539}

\bibitem[\protect\citeauthoryear{{Cantrell}, {Bailyn}, {McClintock}  \&
  {Orosz}}{{Cantrell} et~al.}{2008}]{Cantrell2008}
{Cantrell} A.~G.,  {Bailyn} C.~D.,  {McClintock} J.~E.,   {Orosz} J.~A.,  2008,
  \mn@doi [\apjl] {10.1086/528793}, \href
  {https://ui.adsabs.harvard.edu/abs/2008ApJ...673L.159C} {673, L159}

\bibitem[\protect\citeauthoryear{{Casares}}{{Casares}}{2015}]{Casares2015}
{Casares} J.,  2015, \mn@doi [\apj] {10.1088/0004-637X/808/1/80}, \href
  {https://ui.adsabs.harvard.edu/abs/2015ApJ...808...80C} {808, 80}

\bibitem[\protect\citeauthoryear{{Casares}}{{Casares}}{2016}]{Casares2016}
{Casares} J.,  2016, \mn@doi [\apj] {10.3847/0004-637X/822/2/99}, \href
  {https://ui.adsabs.harvard.edu/abs/2016ApJ...822...99C} {822, 99}

\bibitem[\protect\citeauthoryear{{Casares} \& {Jonker}}{{Casares} \&
  {Jonker}}{2014}]{Casares2014}
{Casares} J.,  {Jonker} P.~G.,  2014, \mn@doi [\ssr]
  {10.1007/s11214-013-0030-6}, \href
  {http://adsabs.harvard.edu/abs/2014SSRv..183..223C} {183, 223}

\bibitem[\protect\citeauthoryear{{Chiba} \& {Beers}}{{Chiba} \&
  {Beers}}{2000}]{Chiba2000}
{Chiba} M.,  {Beers} T.~C.,  2000, \mn@doi [\aj] {10.1086/301409}, \href
  {https://ui.adsabs.harvard.edu/abs/2000AJ....119.2843C} {119, 2843}

\bibitem[\protect\citeauthoryear{{Coelho}}{{Coelho}}{2014}]{Coelho2014}
{Coelho} P.~R.~T.,  2014, \mn@doi [\mnras] {10.1093/mnras/stu365}, \href
  {https://ui.adsabs.harvard.edu/abs/2014MNRAS.440.1027C} {440, 1027}

\bibitem[\protect\citeauthoryear{{Corral-Santana}, {Casares},
  {Mu{\~n}oz-Darias}, {Bauer}, {Mart{\'{\i}}nez-Pais}  \&
  {Russell}}{{Corral-Santana} et~al.}{2016}]{Corral-Santana2016}
{Corral-Santana} J.~M.,  {Casares} J.,  {Mu{\~n}oz-Darias} T.,  {Bauer} F.~E.,
  {Mart{\'{\i}}nez-Pais} I.~G.,   {Russell} D.~M.,  2016, \mn@doi [\aap]
  {10.1051/0004-6361/201527130}, \href
  {https://ui.adsabs.harvard.edu/abs/2016A%26A...587A..61C} {587, A61}

\bibitem[\protect\citeauthoryear{{Corral-Santana} et~al.,}{{Corral-Santana}
  et~al.}{2018}]{Corral-Santana2018}
{Corral-Santana} J.~M.,  et~al., 2018, \mn@doi [\mnras]
  {10.1093/mnras/stx3156}, \href
  {http://adsabs.harvard.edu/abs/2018MNRAS.475.1036C} {475, 1036}

\bibitem[\protect\citeauthoryear{{Deegan}, {Combet}  \& {Wynn}}{{Deegan}
  et~al.}{2009}]{Deegan2009}
{Deegan} P.,  {Combet} C.,   {Wynn} G.~A.,  2009, \mn@doi [\mnras]
  {10.1111/j.1365-2966.2009.15573.x}, \href
  {https://ui.adsabs.harvard.edu/abs/2009MNRAS.400.1337D} {400, 1337}

\bibitem[\protect\citeauthoryear{{Dhawan}, {Mirabel}, {Rib{\'o}}  \&
  {Rodrigues}}{{Dhawan} et~al.}{2007}]{Dhawan2007}
{Dhawan} V.,  {Mirabel} I.~F.,  {Rib{\'o}} M.,   {Rodrigues} I.,  2007, \mn@doi
  [\apj] {10.1086/520111}, \href
  {https://ui.adsabs.harvard.edu/abs/2007ApJ...668..430D} {668, 430}

\bibitem[\protect\citeauthoryear{{Eggleton}}{{Eggleton}}{1983}]{Eggleton1983}
{Eggleton} P.~P.,  1983, \mn@doi [\apj] {10.1086/160960}, \href
  {http://adsabs.harvard.edu/abs/1983ApJ...268..368E} {268, 368}

\bibitem[\protect\citeauthoryear{{Eggleton}, {Faulkner}  \&
  {Cannon}}{{Eggleton} et~al.}{1998}]{Eggleton1998}
{Eggleton} P.~P.,  {Faulkner} J.,   {Cannon} R.~C.,  1998, \mn@doi [\mnras]
  {10.1046/j.1365-8711.1998.01655.x}, \href
  {https://ui.adsabs.harvard.edu/abs/1998MNRAS.298..831E} {298, 831}

\bibitem[\protect\citeauthoryear{{Faulkner}, {Flannery}  \&
  {Warner}}{{Faulkner} et~al.}{1972}]{Faulkner1972}
{Faulkner} J.,  {Flannery} B.~P.,   {Warner} B.,  1972, \mn@doi [\apjl]
  {10.1086/180989}, \href {http://adsabs.harvard.edu/abs/1972ApJ...175L..79F}
  {175, L79}

\bibitem[\protect\citeauthoryear{{Foreman-Mackey}, {Hogg}, {Lang}  \&
  {Goodman}}{{Foreman-Mackey} et~al.}{2013}]{Foreman-Mackey2013}
{Foreman-Mackey} D.,  {Hogg} D.~W.,  {Lang} D.,   {Goodman} J.,  2013, \mn@doi
  [\pasp] {10.1086/670067}, \href
  {https://ui.adsabs.harvard.edu/abs/2013PASP..125..306F} {125, 306}

\bibitem[\protect\citeauthoryear{{Frank}, {King}  \& {Lasota}}{{Frank}
  et~al.}{1987}]{Frank1987}
{Frank} J.,  {King} A.~R.,   {Lasota} J.~P.,  1987, \aap, \href
  {https://ui.adsabs.harvard.edu/abs/1987A&A...178..137F} {178, 137}

\bibitem[\protect\citeauthoryear{{Freudling}, {Romaniello}, {Bramich},
  {Ballester}, {Forchi}, {Garc{\'\i}a-Dabl{\'o}}, {Moehler}  \&
  {Neeser}}{{Freudling} et~al.}{2013}]{Freudling2013}
{Freudling} W.,  {Romaniello} M.,  {Bramich} D.~M.,  {Ballester} P.,  {Forchi}
  V.,  {Garc{\'\i}a-Dabl{\'o}} C.~E.,  {Moehler} S.,   {Neeser} M.~J.,  2013,
  \mn@doi [\aap] {10.1051/0004-6361/201322494}, \href
  {https://ui.adsabs.harvard.edu/abs/2013A&A...559A..96F} {559, A96}

\bibitem[\protect\citeauthoryear{{Gaia Collaboration} et~al.,}{{Gaia
  Collaboration} et~al.}{2016}]{GaiaMission2016}
{Gaia Collaboration} et~al., 2016, \mn@doi [\aap]
  {10.1051/0004-6361/201629272}, \href
  {https://ui.adsabs.harvard.edu/abs/2016A&A...595A...1G} {595, A1}

\bibitem[\protect\citeauthoryear{{Gaia Collaboration} et~al.,}{{Gaia
  Collaboration} et~al.}{2018}]{GaiaDR22018}
{Gaia Collaboration} et~al., 2018, \mn@doi [\aap]
  {10.1051/0004-6361/201833051}, \href
  {https://ui.adsabs.harvard.edu/abs/2018A&A...616A...1G} {616, A1}

\bibitem[\protect\citeauthoryear{{Gaia Collaboration}, {Brown}, {Vallenari},
  {Prusti}, {de Bruijne}, {Babusiaux}  \& {Biermann}}{{Gaia Collaboration}
  et~al.}{2020}]{GaiaEDR32020}
{Gaia Collaboration} {Brown} A.~G.~A.,  {Vallenari} A.,  {Prusti} T.,  {de
  Bruijne} J.~H.~J.,  {Babusiaux} C.,   {Biermann} M.,  2020, arXiv e-prints,
  \href {https://ui.adsabs.harvard.edu/abs/2020arXiv201201533G} {p.
  arXiv:2012.01533}

\bibitem[\protect\citeauthoryear{{Gilmore} \& {Reid}}{{Gilmore} \&
  {Reid}}{1983}]{Gilmore1983}
{Gilmore} G.,  {Reid} N.,  1983, \mnras, \href
  {http://adsabs.harvard.edu/abs/1983MNRAS.202.1025G} {202, 1025}

\bibitem[\protect\citeauthoryear{{Gonz{\'a}lez Hern{\'a}ndez}, {Rebolo},
  {Pe{\~n}arrubia}, {Casares}  \& {Israelian}}{{Gonz{\'a}lez Hern{\'a}ndez}
  et~al.}{2005}]{GonzalezHernandez2005}
{Gonz{\'a}lez Hern{\'a}ndez} J.~I.,  {Rebolo} R.,  {Pe{\~n}arrubia} J.,
  {Casares} J.,   {Israelian} G.,  2005, \mn@doi [\aap]
  {10.1051/0004-6361:20042453}, \href
  {http://adsabs.harvard.edu/abs/2005A%26A...435.1185G} {435, 1185}

\bibitem[\protect\citeauthoryear{{Gray}}{{Gray}}{1992}]{Gray1992}
{Gray} D.~F.,  1992, {The observation and analysis of stellar photospheres.}

\bibitem[\protect\citeauthoryear{{Greiner} et~al.,}{{Greiner}
  et~al.}{2008}]{Greiner2008}
{Greiner} J.,  et~al., 2008, \mn@doi [\pasp] {10.1086/587032}, \href
  {https://ui.adsabs.harvard.edu/abs/2008PASP..120..405G} {120, 405}

\bibitem[\protect\citeauthoryear{{Greiner}, {Rau}  \& {Schady}}{{Greiner}
  et~al.}{2012}]{Greiner2012}
{Greiner} J.,  {Rau} A.,   {Schady} P.,  2012, The Astronomer's Telegram, \href
  {https://ui.adsabs.harvard.edu/abs/2012ATel.4030....1G} {4030, 1}

\bibitem[\protect\citeauthoryear{{Heida}, {Jonker}, {Torres}  \&
  {Chiavassa}}{{Heida} et~al.}{2017}]{Heida2017}
{Heida} M.,  {Jonker} P.~G.,  {Torres} M.~A.~P.,   {Chiavassa} A.,  2017,
  \mn@doi [\apj] {10.3847/1538-4357/aa85df}, \href
  {https://ui.adsabs.harvard.edu/abs/2017ApJ...846..132H} {846, 132}

\bibitem[\protect\citeauthoryear{{Horne} \& {Marsh}}{{Horne} \&
  {Marsh}}{1986}]{Horne1986}
{Horne} K.,  {Marsh} T.~R.,  1986, \mn@doi [\mnras] {10.1093/mnras/218.4.761},
  \href {https://ui.adsabs.harvard.edu/abs/1986MNRAS.218..761H} {218, 761}

\bibitem[\protect\citeauthoryear{{Hynes} \& {Jones}}{{Hynes} \&
  {Jones}}{2009}]{Hynes2009}
{Hynes} R.~I.,  {Jones} E.~D.,  2009, \mn@doi [\apjl]
  {10.1088/0004-637X/697/1/L14}, \href
  {https://ui.adsabs.harvard.edu/abs/2009ApJ...697L..14H} {697, L14}

\bibitem[\protect\citeauthoryear{{Johnson} \& {Soderblom}}{{Johnson} \&
  {Soderblom}}{1987}]{Johnson1987}
{Johnson} D. R.~H.,  {Soderblom} D.~R.,  1987, \mn@doi [\aj] {10.1086/114370},
  \href {https://ui.adsabs.harvard.edu/abs/1987AJ.....93..864J} {93, 864}

\bibitem[\protect\citeauthoryear{{Jordi}, {Grebel}  \& {Ammon}}{{Jordi}
  et~al.}{2006}]{Jordi2006}
{Jordi} K.,  {Grebel} E.~K.,   {Ammon} K.,  2006, \mn@doi [\aap]
  {10.1051/0004-6361:20066082}, \href
  {https://ui.adsabs.harvard.edu/abs/2006A&A...460..339J} {460, 339}

\bibitem[\protect\citeauthoryear{{Kajava}, {Motta}, {Sanna}, {Veledina}, {Del
  Santo}  \& {Segreto}}{{Kajava} et~al.}{2019}]{Kajava2019}
{Kajava} J.~J.~E.,  {Motta} S.~E.,  {Sanna} A.,  {Veledina} A.,  {Del Santo}
  M.,   {Segreto} A.,  2019, \mn@doi [\mnras] {10.1093/mnrasl/slz089}, \href
  {https://ui.adsabs.harvard.edu/abs/2019MNRAS.488L..18K} {488, L18}

\bibitem[\protect\citeauthoryear{{Kalogera} \& {Baym}}{{Kalogera} \&
  {Baym}}{1996}]{Kalogera1996}
{Kalogera} V.,  {Baym} G.,  1996, \mn@doi [\apjl] {10.1086/310296}, \href
  {https://ui.adsabs.harvard.edu/abs/1996ApJ...470L..61K} {470, L61}

\bibitem[\protect\citeauthoryear{{Kennea} et~al.,}{{Kennea}
  et~al.}{2011}]{Kennea2011}
{Kennea} J.~A.,  et~al., 2011, \mn@doi [\apj] {10.1088/0004-637X/736/1/22},
  \href {https://ui.adsabs.harvard.edu/abs/2011ApJ...736...22K} {736, 22}

\bibitem[\protect\citeauthoryear{{Kennea} et~al.,}{{Kennea}
  et~al.}{2012}]{Kennea2012}
{Kennea} J.~A.,  et~al., 2012, The Astronomer's Telegram, \href
  {https://ui.adsabs.harvard.edu/abs/2012ATel.4044....1K} {4044, 1}

\bibitem[\protect\citeauthoryear{{King}}{{King}}{1993}]{King1993}
{King} A.~R.,  1993, \mn@doi [\mnras] {10.1093/mnras/260.1.L5}, \href
  {https://ui.adsabs.harvard.edu/abs/1993MNRAS.260L...5K} {260, L5}

\bibitem[\protect\citeauthoryear{{Kr{\"u}hler} et~al.,}{{Kr{\"u}hler}
  et~al.}{2008}]{Kruehler:2008aa}
{Kr{\"u}hler} T.,  et~al., 2008, \mn@doi [\apj] {10.1086/590240}, \href
  {https://ui.adsabs.harvard.edu/abs/2008ApJ...685..376K} {685, 376}

\bibitem[\protect\citeauthoryear{{Kuulkers} et~al.,}{{Kuulkers}
  et~al.}{2013}]{Kuulkers2013}
{Kuulkers} E.,  et~al., 2013, \mn@doi [\aap] {10.1051/0004-6361/201219447},
  \href {https://ui.adsabs.harvard.edu/abs/2013A%26A...552A..32K} {552, A32}

\bibitem[\protect\citeauthoryear{{Lindegren} et~al.,}{{Lindegren}
  et~al.}{2020}]{GaiaEDR32020b}
{Lindegren} L.,  et~al., 2020, arXiv e-prints, \href
  {https://ui.adsabs.harvard.edu/abs/2020arXiv201203380L} {p. arXiv:2012.03380}

\bibitem[\protect\citeauthoryear{{L{\'o}pez}, {Jonker}, {Torres}, {Heida},
  {Rau}  \& {Steeghs}}{{L{\'o}pez} et~al.}{2019}]{Lopez2019}
{L{\'o}pez} K.~M.,  {Jonker} P.~G.,  {Torres} M.~A.~P.,  {Heida} M.,  {Rau} A.,
    {Steeghs} D.,  2019, \mn@doi [\mnras] {10.1093/mnras/sty2793}, \href
  {https://ui.adsabs.harvard.edu/abs/2019MNRAS.482.2149L} {482, 2149}

\bibitem[\protect\citeauthoryear{{Marsh}, {Horne}  \& {Shipman}}{{Marsh}
  et~al.}{1987}]{Marsh1987}
{Marsh} T.~R.,  {Horne} K.,   {Shipman} H.~L.,  1987, \mnras, \href
  {http://adsabs.harvard.edu/abs/1987MNRAS.225..551M} {225, 551}

\bibitem[\protect\citeauthoryear{{Marsh}, {Robinson}  \& {Wood}}{{Marsh}
  et~al.}{1994}]{Marsh1994}
{Marsh} T.~R.,  {Robinson} E.~L.,   {Wood} J.~H.,  1994, \mnras, \href
  {http://adsabs.harvard.edu/abs/1994MNRAS.266..137M} {266, 137}

\bibitem[\protect\citeauthoryear{{Mata S{\'a}nchez}, {Mu{\~n}oz-Darias},
  {Casares}, {Corral-Santana}  \& {Shahbaz}}{{Mata S{\'a}nchez}
  et~al.}{2015}]{MataSanchez2015b}
{Mata S{\'a}nchez} D.,  {Mu{\~n}oz-Darias} T.,  {Casares} J.,  {Corral-Santana}
  J.~M.,   {Shahbaz} T.,  2015, \mn@doi [\mnras] {10.1093/mnras/stv2111}, \href
  {http://adsabs.harvard.edu/abs/2015MNRAS.454.2199M} {454, 2199}

\bibitem[\protect\citeauthoryear{{Matsuoka} et~al.,}{{Matsuoka}
  et~al.}{2009}]{Matsuoka2009}
{Matsuoka} M.,  et~al., 2009, \mn@doi [\pasj] {10.1093/pasj/61.5.999}, \href
  {https://ui.adsabs.harvard.edu/abs/2009PASJ...61..999M} {61, 999}

\bibitem[\protect\citeauthoryear{{Mihalas} \& {Binney}}{{Mihalas} \&
  {Binney}}{1981}]{Mihalas1981}
{Mihalas} D.,  {Binney} J.,  1981, {Galactic astronomy. Structure and
  kinematics}

\bibitem[\protect\citeauthoryear{{Mihara} et~al.,}{{Mihara}
  et~al.}{2011}]{Mihara2011}
{Mihara} T.,  et~al., 2011, \mn@doi [\pasj] {10.1093/pasj/63.sp3.S623}, \href
  {https://ui.adsabs.harvard.edu/abs/2011PASJ...63S.623M} {63, S623}

\bibitem[\protect\citeauthoryear{{Mirabel}, {Dhawan}, {Mignani}, {Rodrigues}
  \& {Guglielmetti}}{{Mirabel} et~al.}{2001}]{Mirabel2001}
{Mirabel} I.~F.,  {Dhawan} V.,  {Mignani} R.~P.,  {Rodrigues} I.,
  {Guglielmetti} F.,  2001, \mn@doi [\nat] {10.1038/35093060}, \href
  {https://ui.adsabs.harvard.edu/abs/2001Natur.413..139M} {413, 139}

\bibitem[\protect\citeauthoryear{{Morihana} et~al.,}{{Morihana}
  et~al.}{2013}]{Morihana2013}
{Morihana} K.,  et~al., 2013, \mn@doi [\pasj] {10.1093/pasj/65.5.L10}, \href
  {https://ui.adsabs.harvard.edu/abs/2013PASJ...65L..10M} {65, L10}

\bibitem[\protect\citeauthoryear{{Mu{\~n}oz-Darias}, {Torres}  \&
  {Garcia}}{{Mu{\~n}oz-Darias} et~al.}{2018}]{Munoz-Darias2018}
{Mu{\~n}oz-Darias} T.,  {Torres} M. A.~P.,   {Garcia} M.~R.,  2018, \mn@doi
  [\mnras] {10.1093/mnras/sty1711}, \href
  {https://ui.adsabs.harvard.edu/abs/2018MNRAS.479.3987M} {479, 3987}

\bibitem[\protect\citeauthoryear{{Pecaut} \& {Mamajek}}{{Pecaut} \&
  {Mamajek}}{2013}]{Pecaut2013}
{Pecaut} M.~J.,  {Mamajek} E.~E.,  2013, \mn@doi [\apjs]
  {10.1088/0067-0049/208/1/9}, \href
  {https://ui.adsabs.harvard.edu/abs/2013ApJS..208....9P} {208, 9}

\bibitem[\protect\citeauthoryear{{Podsiadlowski}, {Rappaport}  \&
  {Han}}{{Podsiadlowski} et~al.}{2003}]{Podsiadlowski2003}
{Podsiadlowski} P.,  {Rappaport} S.,   {Han} Z.,  2003, \mn@doi [\mnras]
  {10.1046/j.1365-8711.2003.06464.x}, \href
  {https://ui.adsabs.harvard.edu/abs/2003MNRAS.341..385P} {341, 385}

\bibitem[\protect\citeauthoryear{{Predehl} \& {Schmitt}}{{Predehl} \&
  {Schmitt}}{1995}]{Predehl1995}
{Predehl} P.,  {Schmitt} J.~H.~M.~M.,  1995, \aap, \href
  {http://adsabs.harvard.edu/abs/1995A%26A...293..889P} {293}

\bibitem[\protect\citeauthoryear{{Raman}, {Maitra}  \& {Paul}}{{Raman}
  et~al.}{2018}]{Raman2018}
{Raman} G.,  {Maitra} C.,   {Paul} B.,  2018, \mn@doi [\mnras]
  {10.1093/mnras/sty918}, \href
  {https://ui.adsabs.harvard.edu/abs/2018MNRAS.477.5358R} {477, 5358}

\bibitem[\protect\citeauthoryear{{Rappaport} \& {Joss}}{{Rappaport} \&
  {Joss}}{1983}]{Rappaport1983}
{Rappaport} S.~A.,  {Joss} P.~C.,  1983, in {Lewin} W.~H.~G.,  {van den Heuvel}
  E.~P.~J.,  eds, Accretion-Driven Stellar X-ray Sources. p.~13

\bibitem[\protect\citeauthoryear{{Ratti} et~al.,}{{Ratti}
  et~al.}{2012}]{Ratti2012}
{Ratti} E.~M.,  et~al., 2012, \mn@doi [\mnras]
  {10.1111/j.1365-2966.2012.21071.x}, \href
  {https://ui.adsabs.harvard.edu/abs/2012MNRAS.423.2656R} {423, 2656}

\bibitem[\protect\citeauthoryear{{Rodr{\'\i}guez-Gil}
  et~al.,}{{Rodr{\'\i}guez-Gil} et~al.}{2015}]{RodriguezGil2015}
{Rodr{\'\i}guez-Gil} P.,  et~al., 2015, \mn@doi [\mnras]
  {10.1093/mnras/stv1244}, \href
  {https://ui.adsabs.harvard.edu/abs/2015MNRAS.452..146R} {452, 146}

\bibitem[\protect\citeauthoryear{{Sato} et~al.,}{{Sato}
  et~al.}{2012}]{Sato2012}
{Sato} R.,  et~al., 2012, The Astronomer's Telegram, \href
  {https://ui.adsabs.harvard.edu/abs/2012ATel.4024....1S} {4024}

\bibitem[\protect\citeauthoryear{{Schlafly} \& {Finkbeiner}}{{Schlafly} \&
  {Finkbeiner}}{2011}]{Schlafly2011}
{Schlafly} E.~F.,  {Finkbeiner} D.~P.,  2011, \mn@doi [\apj]
  {10.1088/0004-637X/737/2/103}, \href
  {https://ui.adsabs.harvard.edu/abs/2011ApJ...737..103S} {737, 103}

\bibitem[\protect\citeauthoryear{{Schoembs} \& {Hartmann}}{{Schoembs} \&
  {Hartmann}}{1983}]{Schoembs1983}
{Schoembs} R.,  {Hartmann} K.,  1983, \aap, \href
  {http://adsabs.harvard.edu/abs/1983A%26A...128...37S} {128, 37}

\bibitem[\protect\citeauthoryear{{Sch{\"o}nberg} \&
  {Chandrasekhar}}{{Sch{\"o}nberg} \& {Chandrasekhar}}{1942}]{SC1942}
{Sch{\"o}nberg} M.,  {Chandrasekhar} S.,  1942, \mn@doi [\apj]
  {10.1086/144444}, \href
  {https://ui.adsabs.harvard.edu/abs/1942ApJ....96..161S} {96, 161}

\bibitem[\protect\citeauthoryear{{Shafter}, {Szkody}  \&
  {Thorstensen}}{{Shafter} et~al.}{1986}]{Shafter1986}
{Shafter} A.~W.,  {Szkody} P.,   {Thorstensen} J.~R.,  1986, \mn@doi [\apj]
  {10.1086/164549}, \href
  {https://ui.adsabs.harvard.edu/abs/1986ApJ...308..765S} {308, 765}

\bibitem[\protect\citeauthoryear{{Shaw}, {Charles}, {Casares}  \&
  {Steeghs}}{{Shaw} et~al.}{2017}]{Shaw2017}
{Shaw} A.~W.,  {Charles} P.~A.,  {Casares} J.,   {Steeghs} D.,  2017, in 7
  years of MAXI: monitoring X-ray Transients. p.~45

\bibitem[\protect\citeauthoryear{{Shidatsu} et~al.,}{{Shidatsu}
  et~al.}{2013}]{Shidatsu2013}
{Shidatsu} M.,  et~al., 2013, \mn@doi [\apj] {10.1088/0004-637X/779/1/26},
  \href {https://ui.adsabs.harvard.edu/abs/2013ApJ...779...26S} {779, 26}

\bibitem[\protect\citeauthoryear{{Soubiran}, {Bienaym{\'e}}  \&
  {Siebert}}{{Soubiran} et~al.}{2003}]{Soubiran2003}
{Soubiran} C.,  {Bienaym{\'e}} O.,   {Siebert} A.,  2003, \mn@doi [\aap]
  {10.1051/0004-6361:20021615}, \href
  {https://ui.adsabs.harvard.edu/abs/2003A&A...398..141S} {398, 141}

\bibitem[\protect\citeauthoryear{{Steeghs} \& {Jonker}}{{Steeghs} \&
  {Jonker}}{2007}]{Jonker2007}
{Steeghs} D.,  {Jonker} P.~G.,  2007, \mn@doi [\apjl] {10.1086/523848}, \href
  {https://ui.adsabs.harvard.edu/abs/2007ApJ...669L..85S} {669, L85}

\bibitem[\protect\citeauthoryear{{Suwa} et~al.,}{{Suwa}
  et~al.}{2012}]{Suwa2012}
{Suwa} F.,  et~al., 2012, The Astronomer's Telegram, \href
  {https://ui.adsabs.harvard.edu/abs/2012ATel.4035....1S} {4035, 1}

\bibitem[\protect\citeauthoryear{{Torres}, {Jonker}, {Miller-Jones}, {Steeghs},
  {Repetto}  \& {Wu}}{{Torres} et~al.}{2015}]{Torres2015}
{Torres} M.~A.~P.,  {Jonker} P.~G.,  {Miller-Jones} J.~C.~A.,  {Steeghs} D.,
  {Repetto} S.,   {Wu} J.,  2015, \mn@doi [\mnras] {10.1093/mnras/stv720},
  \href {http://adsabs.harvard.edu/abs/2015MNRAS.450.4292T} {450, 4292}

\bibitem[\protect\citeauthoryear{{Torres} et~al.,}{{Torres}
  et~al.}{2019a}]{Torres2019a}
{Torres} M.~A.~P.,  et~al., 2019a, \mn@doi [\mnras] {10.1093/mnras/stz1405},
  \href {https://ui.adsabs.harvard.edu/abs/2019MNRAS.487.2296T} {487, 2296}

\bibitem[\protect\citeauthoryear{{Torres}, {Casares}, {Jim{\'e}nez-Ibarra},
  {Mu{\~n}oz-Darias}, {Armas Padilla}, {Jonker}  \& {Heida}}{{Torres}
  et~al.}{2019b}]{Torres2019}
{Torres} M.~A.~P.,  {Casares} J.,  {Jim{\'e}nez-Ibarra} F.,  {Mu{\~n}oz-Darias}
  T.,  {Armas Padilla} M.,  {Jonker} P.~G.,   {Heida} M.,  2019b, \mn@doi
  [\apjl] {10.3847/2041-8213/ab39df}, \href
  {https://ui.adsabs.harvard.edu/abs/2019ApJ...882L..21T} {882, L21}

\bibitem[\protect\citeauthoryear{{Torres}, {Casares}, {Jim{\'e}nez-Ibarra},
  {{\'A}lvarez-Hern{\'a}ndez}, {Mu{\~n}oz-Darias}, {Armas Padilla}, {Jonker}
  \& {Heida}}{{Torres} et~al.}{2020}]{Torres2020}
{Torres} M.~A.~P.,  {Casares} J.,  {Jim{\'e}nez-Ibarra} F.,
  {{\'A}lvarez-Hern{\'a}ndez} A.,  {Mu{\~n}oz-Darias} T.,  {Armas Padilla} M.,
  {Jonker} P.~G.,   {Heida} M.,  2020, \mn@doi [\apjl]
  {10.3847/2041-8213/ab863a}, \href
  {https://ui.adsabs.harvard.edu/abs/2020ApJ...893L..37T} {893, L37}

\bibitem[\protect\citeauthoryear{{Torres}, {Jonker}, {Casares}, {Miller-Jones}
  \& {Steeghs}}{{Torres} et~al.}{2021}]{Torres2021}
{Torres} M.~A.~P.,  {Jonker} P.~G.,  {Casares} J.,  {Miller-Jones} J.~C.~A.,
  {Steeghs} D.,  2021, \mn@doi [\mnras] {10.1093/mnras/staa3786}, \href
  {https://ui.adsabs.harvard.edu/abs/2021MNRAS.501.2174T} {501, 2174}

\bibitem[\protect\citeauthoryear{{Valdes}, {Gupta}, {Rose}, {Singh}  \&
  {Bell}}{{Valdes} et~al.}{2004}]{Valdes2004}
{Valdes} F.,  {Gupta} R.,  {Rose} J.~A.,  {Singh} H.~P.,   {Bell} D.~J.,  2004,
  \mn@doi [\apjs] {10.1086/386343}, \href
  {http://adsabs.harvard.edu/abs/2004ApJS..152..251V} {152, 251}

\bibitem[\protect\citeauthoryear{{Wade} \& {Horne}}{{Wade} \&
  {Horne}}{1988}]{Wade1988}
{Wade} R.~A.,  {Horne} K.,  1988, \mn@doi [\apj] {10.1086/165905}, \href
  {http://adsabs.harvard.edu/abs/1988ApJ...324..411W} {324, 411}

\bibitem[\protect\citeauthoryear{{White} \& {Swank}}{{White} \&
  {Swank}}{1982}]{White1982}
{White} N.~E.,  {Swank} J.~H.,  1982, \mn@doi [\apjl] {10.1086/183737}, \href
  {https://ui.adsabs.harvard.edu/abs/1982ApJ...253L..61W} {253, L61}

\bibitem[\protect\citeauthoryear{{Whitehurst} \& {King}}{{Whitehurst} \&
  {King}}{1991}]{Whitehurst1991}
{Whitehurst} R.,  {King} A.,  1991, \mn@doi [\mnras] {10.1093/mnras/249.1.25},
  \href {https://ui.adsabs.harvard.edu/abs/1991MNRAS.249...25W} {249, 25}

\bibitem[\protect\citeauthoryear{{Willems}, {Henninger}, {Levin}, {Ivanova},
  {Kalogera}, {McGhee}, {Timmes}  \& {Fryer}}{{Willems}
  et~al.}{2005}]{Willems2005}
{Willems} B.,  {Henninger} M.,  {Levin} T.,  {Ivanova} N.,  {Kalogera} V.,
  {McGhee} K.,  {Timmes} F.~X.,   {Fryer} C.~L.,  2005, \mn@doi [\apj]
  {10.1086/429557}, \href
  {https://ui.adsabs.harvard.edu/abs/2005ApJ...625..324W} {625, 324}

\bibitem[\protect\citeauthoryear{{Zi{\'o}{\l}kowski} \&
  {Zdziarski}}{{Zi{\'o}{\l}kowski} \& {Zdziarski}}{2020}]{Ziolkowski2020}
{Zi{\'o}{\l}kowski} J.,  {Zdziarski} A.~A.,  2020, \mn@doi [\mnras]
  {10.1093/mnras/staa3088}, \href
  {https://ui.adsabs.harvard.edu/abs/2020MNRAS.499.4832Z} {499, 4832}

\bibitem[\protect\citeauthoryear{{van Grunsven}, {Jonker}, {Verbunt}  \&
  {Robinson}}{{van Grunsven} et~al.}{2017}]{vanGrunsven2017}
{van Grunsven} T. F.~J.,  {Jonker} P.~G.,  {Verbunt} F. W.~M.,   {Robinson}
  E.~L.,  2017, \mn@doi [\mnras] {10.1093/mnras/stx2071}, \href
  {https://ui.adsabs.harvard.edu/abs/2017MNRAS.472.1907V} {472, 1907}

\makeatother
\end{thebibliography}
